\documentclass[aps,pre,superscriptaddress,twocolumn,balancelastpage]{revtex4-1}

\usepackage{times}
\usepackage{amsmath,amssymb}
\usepackage{graphicx}
\graphicspath{{figures/}{/}}
\usepackage{verbatim}
\usepackage{color}

\graphicspath{{/}{figures/}}

\newcommand{\eqn}  {Eq.~}
\newcommand{\eqns} {Eqs.~}

\newcommand  {\ee}            {\mathrm{e}}
\newcommand  {\ii}            {\mathrm{i}}
\newcommand  {\nup}           {N_{\mathord{\uparrow}}}
\newcommand  {\nb}            {N_\mathrm{b}}
\newcommand  {\nf}            {N_\mathrm{f}}
\newcommand  {\nn}            {N}

\newcommand  {\frdim}         {\mathcal{D}}
\newcommand  {\frdimq}        {\frdim_q}

\newcommand  {\ket}[1]        {\lvert#1\rangle}
\newcommand  {\bra}[1]        {\langle#1\rvert}
\newcommand  {\braket}[2]     {\langle#1\vert#2\rangle}

\newcommand  {\abs}[1]        {\lvert#1\rvert}

\newcommand  {\psicircle}     {\psi^{\circ}}
\newcommand  {\psisquare}     {\psi^{\Box}}

\newcommand  {\per}           {\mathop{\mathrm{per}}\nolimits}
\newcommand  {\Tr}            {\mathop{\mathrm{Tr}}\nolimits}

\newcommand  {\TrB}           {\mathop{\Tr_\mathrm{B}}\nolimits}
\newcommand  {\rhoA}          {\rho_\mathrm{A}}

\makeatletter
\let\Hy@backout\@gobble
\makeatother

\begin{document}

\title{Statistical properties of eigenstate amplitudes in complex quantum systems}

\author{Wouter Beugeling}
\affiliation{Max-Planck-Institut f\"ur Physik komplexer Systeme, N\"othnitzer Stra\ss e 38, 01187 Dresden, Germany}
\affiliation{Lehrstuhl f\"ur Theoretische Physik I \& II, Technische Universit\"at Dortmund, Otto-Hahn-Stra\ss e 4, 44221 Dortmund, Germany}
\affiliation{Physikalisches Institut, Universit\"at W\"urzburg, Am Hubland, 97074 W\"urzburg, Germany}

\author{Arnd B\"acker}
\affiliation{Max-Planck-Institut f\"ur Physik komplexer Systeme, N\"othnitzer Stra\ss e 38, 01187 Dresden, Germany}
\affiliation{Technische Universit\"at Dresden, Institut f\"ur Theoretische Physik and Center for Dynamics, 01062 Dresden, Germany}

\author{Roderich Moessner}
\affiliation{Max-Planck-Institut f\"ur Physik komplexer Systeme, N\"othnitzer Stra\ss e 38, 01187 Dresden, Germany}

\author{Masudul Haque}
\affiliation{Max-Planck-Institut f\"ur Physik komplexer Systeme, N\"othnitzer Stra\ss e 38, 01187
  Dresden, Germany}
\affiliation{Department of Theoretical Physics, Maynooth University, Co. Kildare, Ireland}

\date{\today}
\pacs{}

\begin{abstract}

  We study the eigenstates of quantum systems with large Hilbert spaces, via their distribution of
  wavefunction amplitudes in a real-space basis.  For single-particle `quantum billiards', these
  real-space amplitudes are known to have Gaussian distribution for chaotic systems.  In this work,
  we formulate and address the corresponding question for many-body lattice quantum systems.  For
  integrable many-body systems, we examine the deviation from Gaussianity and provide evidence that
  the distribution generically tends toward power-law behavior in the limit of large sizes.  We
  relate the deviation from Gaussianity to the entanglement content of many-body eigenstates.  For
  integrable billiards, we find several cases where the distribution has power-law tails.

\end{abstract}

\maketitle


\section{Introduction}

Except for particularly simple systems, eigenstates of quantum Hamiltonians are complex objects,
described by an exponentially large number of coefficients (amplitudes).  Energy eigenstates are
constitutive to the formulation of quantum mechanics.  They are also essential in describing closed
quantum systems, e.g., in considerations of thermalization \cite{Deutsch1991, Srednicki1994,
  RigolEA2008, PolkovnikovRigol_AdvPhys2016, BorgonoviIzrailevSantos_PhysRep2016}.  Thus, one might
reasonably regard the structure of eigenstates, e.g., the statistical properties of amplitudes, as
being fundamental to our understanding of the quantum world.  Amplitude distributions have 
been studied for single-particle (quantum billiard) systems \cite{ShapiroGoelman_PRL1984,
  McDonaldKaufman_PRA1988, AurichSteiner_PhysicaD1991, AurichSteiner_PhysicaD1993, LiRobnik_JPA1994,
  SimmmelEckert_PhysicaD1996, Prosen_PhysLett97, Backer2003, Backer2007}.  However, little is known
about corresponding distributions for quantum many-body Hamiltonians.  In this work we address
distributions of coefficients (in the basis of real-space configurations), clarifying in particular
the consequences of \emph{integrability}.

While it is difficult to find a universally accepted definition of quantum integrability
\cite{book_Sutherland_Beautiful_2004, CauxMossel_Integrability_JSTAT2011}, we will refer to systems
with Poissonian level spacing statistics (within a single symmetry sector) to be integrable or
regular, and to those with random-matrix statistics as non-integrable or chaotic.  This distinction
appears both in single-particle billiards \cite{BohigasEA_PRL1984, BerryTabor_ProcRoy1977} and in
many-body systems \cite{Montambaux_Poilblanc_PRL1993, Poilblanc_Ziman_Mila_Montambaux_EPL1993,
  HsuEA_PRB1993, NarozhnyMillis_PRB2004, Kolovsky_Buchleitner_EPL2004, KudoDeguchi_JPSJ2005,
  KarthikEA_PRA2007, Santos_JMathPhys2009, KollathEA2010, SantosRigol2010a, SantosEA2012,
  BogomolnyRoux_PRL2014}.  This operational definition is inadequate in some situations, but will
suffice for this work.  Many-body integrable systems include non-interacting (`free') fermions, free
bosons, and systems solvable by Bethe ansatz.  Integrable quantum billiards are those whose
corresponding classical problems have as many independent conserved quantities as degrees of
freedom.
Some further comments on integrability are provided in Appendix \ref{suppsec_integrability}.

For quantum billiard systems, the distribution of real-space amplitudes $\psi(\vec{x}) =
\braket{\vec{x}}{\psi}$ of eigenstates has been studied both for chaotic and for mixed systems
\cite{ShapiroGoelman_PRL1984, McDonaldKaufman_PRA1988, AurichSteiner_PhysicaD1991,
  AurichSteiner_PhysicaD1993, LiRobnik_JPA1994, SimmmelEckert_PhysicaD1996, Prosen_PhysLett97,
  BaeckerSchubert_JPA2002, Backer2007}.
In the chaotic case the amplitude distribution is expected to be Gaussian for almost all eigenstates
(with some possible exceptions \cite{Hel1984, BaeSchSti1997}).
This follows from the conjecture that high-energy eigenstates of
chaotic billiards resemble random superpositions of many plane waves leading to a Gaussian
distribution by the central limit theorem \cite{Ber1977b,McDonaldKaufman_PRA1988}.
For single-particle systems, the particle position is the natural basis in which to express the
amplitudes.
In the many-body case, the choice of basis is less obvious, but a direct generalization
is the basis of many-body configurations in real space.  For lattice systems, this is also a widely
used basis for numerical diagonalization.  Our study focuses on coefficients in this basis.

For \emph{non-integrable} systems, we show that eigenstates away from spectral edges have Gaussian
coefficient distributions.  The resemblance to Gaussian form improves with increasing deviation from
integrability, and also improves systematically with system size.
For \emph{integrable} many-body systems, there is clear deviation from Gaussian shape.  We provide
evidence that the distribution approaches a power law as the size is increased.  The convergence is
extremely slow --- a meaningful scaling analysis could only be performed for free fermions, but data for
several integrable systems show the same trend.  An analytic argument is constructed for a toy model
of distinguishable particles, which accounts for the power-law form and the slow convergence.  The
presented numerical data and arguments, taken together, naturally lead to the conjecture that
eigenstates of integrable many-body systems generically have power-law coefficient distributions in
the large-size (`thermodynamic') limit.  This conjecture is remarkable because `generic' results are
usually expected for chaotic rather than integrable systems.

We relate the coefficient distribution to the \emph{entanglement entropy} between two spatial
partitions.  We show that larger deviations from Gaussian shape correlate strongly with low
entanglement, and provide intuition for this correlation.

We also present some results for integrable quantum billiard systems.  Explicit calculation shows in
a few cases that the amplitude distributions have power-law tails.  An extended power-law region can
appear when the regular eigenfunctions contain many inequivalent peaks.  The feature is intriguing
but is not present in all integrable billiard systems.

This article is structured as follows.  In Sec.~\ref{sec_many_body}, we introduce the many-body
models and present a general study of their coefficient distributions, highlighting the differences
between non-integrable and integrable many-body systems, the deviation from Gaussianity and the
correlation of this deviation with the entanglement entropy.  In Sec.~\ref{sec_integrable}, we focus
on integrable many-body systems, and provide numerical evidence and argumentation supporting
approach to power-law behavior in the large-size limit.  In Sec.~\ref{sec_billiards} we consider
several single-particle quantum billiard systems and present results on amplitude distributions for
several integrable and weakly non-integrable billiards.  Section~\ref{sec_conclusion} discusses the
context and some implications of our results.  We provide additional data and supporting discussions
in the appendices.


\section{Many-body quantum systems}
\label{sec_many_body}

In this section, we introduce the many-body Hamiltonians we use in this paper
(Sec.\ \ref{subsec_models}), and then present a general overview of the coefficient distributions
(Sec.\ \ref{subsec_distns_overview}).  The distributions are close to Gaussian away from the
spectral edges in non-integrable systems.  They deviate significantly from Gaussianity for many
eigenstates in integrable systems, and for eigenstates at the spectral edges in all systems.  In
Sec.\ \ref{subsec_deviation_KLD} we quantify the deviation from Gaussianity using the
Kullback-Leibler divergence, and investigate the degree of Gaussianity in various cases using this
measure.  We show that, with increasing system size, resemblance to Gaussian form improves for
non-integrable systems but deteriorates for integrable systems.

\subsection{Models}
\label{subsec_models}

We consider the spin-$\frac{1}{2}$ XXZ and Bose-Hubbard systems, on finite one-dimensional chains.
We use open boundary conditions to avoid complications due to translation symmetry.  For the XXZ
chain, a next-nearest-neighbor (NNN) coupling breaks integrability:
\begin{equation}\label{eqn_hxxz}
  H_\mathrm{XXZ} = \sum_{i=1}^{L-1} h_{i,i+1} + \lambda \sum_{i=2}^{L-2} h_{i,i+2},
\end{equation}
where $h_{i,j}= S^x_iS^x_j+S^y_iS^y_j +\Delta S^z_iS^z_j$ (with spin-$\tfrac{1}{2}$ operators
$S^{x,y,z}_i$) and $L$ is the number of sites.  The NNN (second) term excludes the coupling between
sites $1$ and $3$, breaking reflection symmetry for $\lambda\not=0$.  We use $\Delta=0.8$ throughout this work. 

The Bose-Hubbard chain is described by the Hamiltonian
\begin{equation}\label{eqn_hbh}
  H_\mathrm{BH} = \sum_{i=1}^{L-1} (b_i^\dagger b_{i+1}+ b_{i+1}^\dagger b_{i}) + \lambda \sum_{i=1}^{L} b_i^\dagger b_i^\dagger b_i b_i,
\end{equation}
where $b_i$ denotes the bosonic annihilation operator on site $i$.

The number $\nup$ of ``up'' spins (XXZ) and the number of bosons $\nb$ (Bose-Hubbard) are conserved
quantities.  We study a single sector at a time, i.e., we fix $(L,\nup)$ or $(L,\nb)$.
The Hilbert space dimensions are $D=\binom{L}{\nup}$ for the XXZ chain and $D=\binom{L+\nb-1}{\nb}$
for the Bose-Hubbard system.

In both Hamiltonians \eqref{eqn_hxxz} and \eqref{eqn_hbh}, the second term breaks integrability; the
dimensionless parameter $\lambda$ controls proximity to integrability.  The two integrable
($\lambda=0$) Hamiltonians are the nearest-neighbor XXZ Hamiltonian, which is integrable by Bethe
ansatz, and a chain of free (non-interacting) bosons, which is integrable due to the absence of
interactions.  
We will present data mostly for $\lambda=1$ (typical non-integrable case) and $\lambda=0$
(integrable case).

In addition to the Hamiltonians classes \eqref{eqn_hxxz} and \eqref{eqn_hbh}, for our detailed
treatment of integrable systems we will also consider a tight-binding system of $\nf$ free fermions
on an $L$-site chain,
\begin{equation}\label{eqn_hff}
H_{\textrm{FF}}= \sum_{i}
(c_i^\dagger c_{i+1}+ c_{i+1}^\dagger c_{i}) + \sum_{i} V_i c_i^\dagger c_{i},
\end{equation}
subject to a weakly varying potential $V_i$ which leaves the system integrable but avoids lattice
symmetries.  Here $c_i$ denotes the fermionic annihilation operator on site $i$.  The Hamiltonian
conserves the fermion number $\nf$.  The Hilbert space dimension is $D=\binom{L}{\nf}$.

\begin{figure}
  \center%
  \includegraphics[width=85mm]{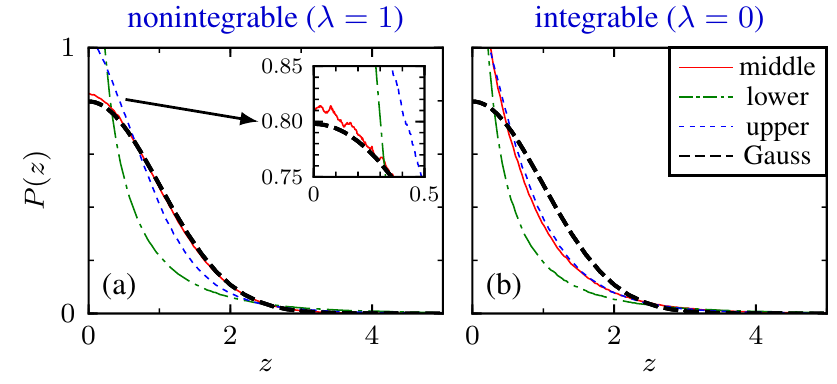}
  \caption{(a) Amplitude distributions for non-integrable XXZ chain, with $(L,\nup)=(17,8)$,  $\Delta=0.8$, and $\lambda=1$. The distributions are over $250$ eigenstates in the middle
    ($E\approx0$), at the lower edge, and at the upper edge of the spectrum. The black dashed curve
    is the Gaussian distribution with unit variance. The inset shows a magnification near zero. (b)
    Same, for the integrable XXZ chain, with $\lambda = 0$.}
\label{fig_manybody_coeff_dist}
\end{figure}

\subsection{Gaussian and non-Gaussian distributions}
\label{subsec_distns_overview}

We are interested in the statistics of coefficients
$c^{(\alpha)}_\gamma\equiv\braket{\phi_\gamma}{\psi_\alpha}$ of the energy eigenstates
$\ket{\psi_\alpha}$.  The basis states $\{\ket{\phi_\gamma}\}$ are spatial configurations, i.e.,
eigenstates of the local operators $S^z_i$ or $b_i^\dagger b_i$.
Normalization ensures that $\sum_{\gamma=1}^{D}\abs{c^{(\alpha)}_\gamma}^2=1$.
We study distributions of $z = \abs{c_\gamma}\sqrt{D}$ (eigenstate indices $\alpha$ are suppressed).
These distributions $P(z)$ then have unit variance, which simplifies comparison between different
sizes.

In Fig.~\ref{fig_manybody_coeff_dist}, we show the distributions of coefficients of 250 eigenstates,
taken from the edges and from the middle of the spectra.
States at the edge are special; they tend to be non-generic (`integrable-like').  In the
coefficients this is manifested by non-Gaussian distributions, regardless of whether the system is
integrable or not.
For integrable many-body systems, e.g., the XXZ chain with $\lambda=0$
[Fig.~\ref{fig_manybody_coeff_dist}(b)] and other cases shown later, the distribution is also
markedly non-Gaussian for eigenstates in the middle of the spectrum.

In contrast to the cases discussed above, for non-integrable systems (e.g., the XXZ chain with NNN coupling at $\lambda=1$), the distribution of the coefficients of the eigenstate in the middle of the spectrum has overall
Gaussian behavior, see Fig.~\ref{fig_manybody_coeff_dist}(a).  The same is valid for the Bose-Hubbard
chain with $\lambda\sim1$ (not shown).
A Gaussian $P(z)$ is expected for complex non-integrable Hamiltonians --- it is equivalent to the
Porter-Thomas distribution for $|z|^2$ in nuclear physics \cite{PorterThomas_PR1956}, and has been
assumed or tested for condensed-matter Hamiltonians, e.g., in Ref.~\cite{LuitzBarlev_PRL16,
  MondainiRigol_PRE2017, MondainiSrednickiRigol_PRE16, AtasBogomolny_JPhysA2017}.

We observe a weak deviation from the Gaussian close to zero (Fig.~\ref{fig_manybody_coeff_dist},
inset)
The small excess weight near $z=0$ is balanced at intermediate values of $z$---the distribution is
lower than the Gaussian at intermediate $z$ and then overshoots the Gaussian curve again at large
$z$.  The overall distribution thus has higher kurtosis than the Gaussian---about $3.17$ for the
data shown in Fig.~\ref{fig_manybody_coeff_dist}(a).
The deviation is characterized in some detail in the following subsection.
%

\begin{figure}
  \center%
  \includegraphics[width=85mm]{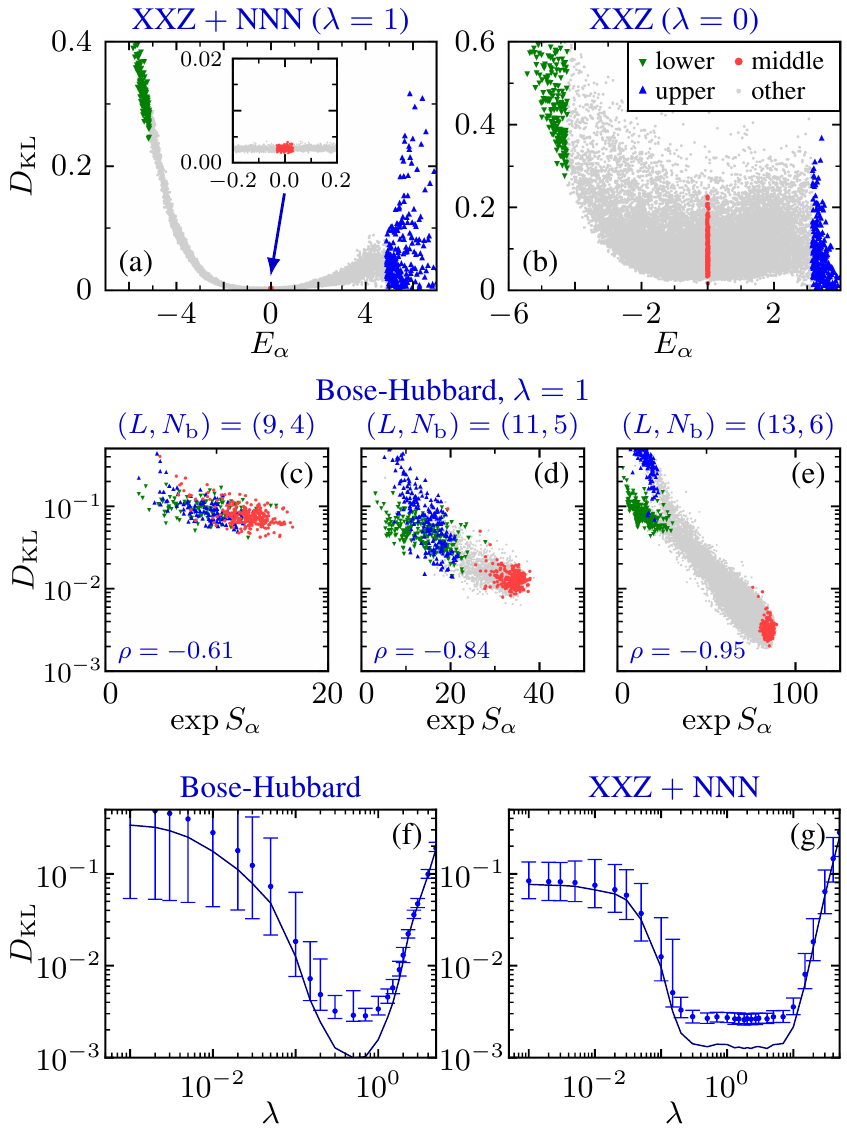}
  \caption{Kullback-Leibler divergence $D_\mathrm{KL}$ (deviation from Gaussianity),
    using $(L,\nup)=(17,8)$ for XXZ and $(L,\nb)=(13,6)$ for Bose-Hubbard. (a) Per-eigenstate
    $D_\mathrm{KL}$ versus eigenenergy $E_\alpha$ for the XXZ chain ($\Delta=0.8$) with and without NNN coupling.  The highlighted states correspond to distributions shown in
    Fig.~\ref{fig_manybody_coeff_dist}.  (c)--(e) Per-eigenstate $D_\mathrm{KL}$ against
    exponentiated entanglement entropy for three system sizes of the Bose-Hubbard model. We indicate
    the Pearson correlation coefficient  $\rho$.  (f,g) $D_\mathrm{KL}$ as function of $\lambda$.
    Data points and error bars are average and standard deviation of the per-eigenstate
    $D_\mathrm{KL}$ values for $250$ eigenstates in the middle of the spectrum.  Solid line is
    $D_\mathrm{KL}$ of the distribution of all coefficients of these states together.  }
\label{fig_dkl_main}
\end{figure}

\subsection{Deviation from Gaussianity}
\label{subsec_deviation_KLD}

We will now present a quantitative analysis of deviation from ``Gaussianity''.  For this purpose we
use a commonly used measure of the difference between two distributions, namely the Kullback-Leibler
divergence (KLD) \cite{KullbackLeibler1951}.  The KLD between $P(z)$ and the standard Gaussian
distribution $P^\mathrm{G}(z)$ is 
\begin{equation}\label{eqn_kld}
  D_{\mathrm{KL}}(P\|P^\mathrm{G}) = \int_0^\infty P(z)\log\frac{P(z)}{P^\mathrm{G}(z)}dz. 
\end{equation}
This quantity vanishes if $P(z)$ is identical to $P^\mathrm{G}(z)$ and grows as $P(z)$ increasingly
deviates from $P^\mathrm{G}(z)$.

In Figs.~\ref{fig_dkl_main}(a,b), the KLD for each eigenstate is plotted against corresponding
eigenenergies.
Consistent with Fig.~\ref{fig_manybody_coeff_dist}(a), in the non-integrable case [Fig.~\ref{fig_dkl_main}(a)], the
$D_{\mathrm{KL}}$ values are close to zero in the middle and larger at the edges of the spectrum.
In the integrable case [Fig.~\ref{fig_dkl_main}(b)], there is a large spread of $D_{\mathrm{KL}}$ throughout the spectrum.

This behavior is reminiscent of that of bipartite entanglement entropy (EE) $S_\alpha$ of
eigenstates
\footnote{Here, we separate the system in two spatial parts A and B of (nearly) equal size.
Given a state $\protect{\ket{\psi}}$, the entanglement entropy is $S=-\Tr \rhoA \log \rhoA$, where
$\rhoA=\TrB \protect{\ket{\psi}}\protect{\bra{\psi}}$ is the reduced density matrix. The spatial
partition is most relevant to the present study of amplitudes in the basis of real-space
configurations.}:
in integrable systems,
the middle of the spectrum has both generic, high-EE eigenstates but also a substantial number of
non-generic, low-EE eigenstates \cite{BeugelingEA2015JStatMech, alba2009entanglement}, while
non-integrable systems have only high-EE eigenstates in the middle of the spectrum
\cite{BeugelingEA2015JStatMech, GarrisonGrover2018PRX, VidmarRigol2017_entanglement}.  This
suggests that the KLD of an eigenstate is inversely correlated with EE, and that large KLD and small
EE both represent deviations from generic (effectively-random or `thermal') behavior.

The correlation between the KLD and EE is shown in Fig.~\ref{fig_dkl_main}(c)--(e), using scatter
plots of the per-eigenstate KLD against the per-eigenstate exponentiated EE,
$\exp(S_\mathrm{\alpha})$.  Here the entanglement is between two spatially connected parts of the
Bose-Hubbard chain, of sizes $l$ and $l+1$, where $2l+1=L$.  The data exhibits a very significant correlation
between the KLD and EE, with improving correlation for increasing system size.  We quantify this
correlation using Pearson correlation coefficients $\rho$ between $\exp S_\alpha$ and $\log
D_\mathrm{KL}$, which measures how linear the correlation between these two quantities is.
The coefficient is negative because larger-KLD states generally have smaller entanglement, i.e., the
plots overall have negative slope.  The magnitude of $\rho$ increases steadily
with system size.  Similar behavior is observed for the XXZ model with NNN couplings (not shown),
which suggests that the improvement of correlation with increasing system size is a generic feature.

The participation ratio (PR) of eigenstates in the real-space configuration basis is more directly
correlated with the KLD.  The inverse PR
\begin{equation}\label{eqn_inverse_pr}
  p^{-1} = D\sum_\gamma \abs{c_\gamma}^4 = \int z^4 P(z) dz
\end{equation}
is the kurtosis of the coefficient distribution, having the value $p=1/3$ for a Gaussian
distribution.  The (inverse) PR has been used as a characterization of proximity to integrability
\cite{BorgonoviIzrailevSantos_PhysRep2016, BeugelingEA2015JStatMech, SantosRigol2010,
  SantosRigol2010a}.

Like the KLD, the PR can be calculated directly from the shape of the coefficient distribution; in
contrast, the EE involves a partial trace which requires additional information about the spatial
structure of the basis states.  In view of the correlation between EE and KLD displayed in
Figs.~\ref{fig_dkl_main}(c)--(e), it is thus expected that the EE and the PR should be
positively correlated, as explored in Ref.\ \cite{BeugelingEA2015JStatMech}.  We provide some
further data in Appendix \ref{app_EE_PR}.

Figures~\ref{fig_dkl_main}(f,g) show the KLD as a function of the integrability-breaking parameter
$\lambda$.  In the non-integrable regime ($\lambda\sim1$) the coefficient distribution for every
eigenstate in the middle of the spectrum is close to Gaussian, with $D_{\mathrm{KL}}$ near zero.
For $\lambda\to 0$, the values of $D_\mathrm{KL}$ grow, and there is a large variation between the
different eigenstates, reflecting the large spread of $D_\mathrm{KL}$ values in
Fig.~\ref{fig_dkl_main}(b).  For $\lambda\gg1$, $D_\mathrm{KL}$ increases rapidly.  In this limit,
local conserved quantities divide the Hilbert space into uncoupled sectors, leading to a large
number of zero coefficients, which accounts for the strong deviation from Gaussianity.

\begin{figure}
  \center%
  \includegraphics[width=85mm]{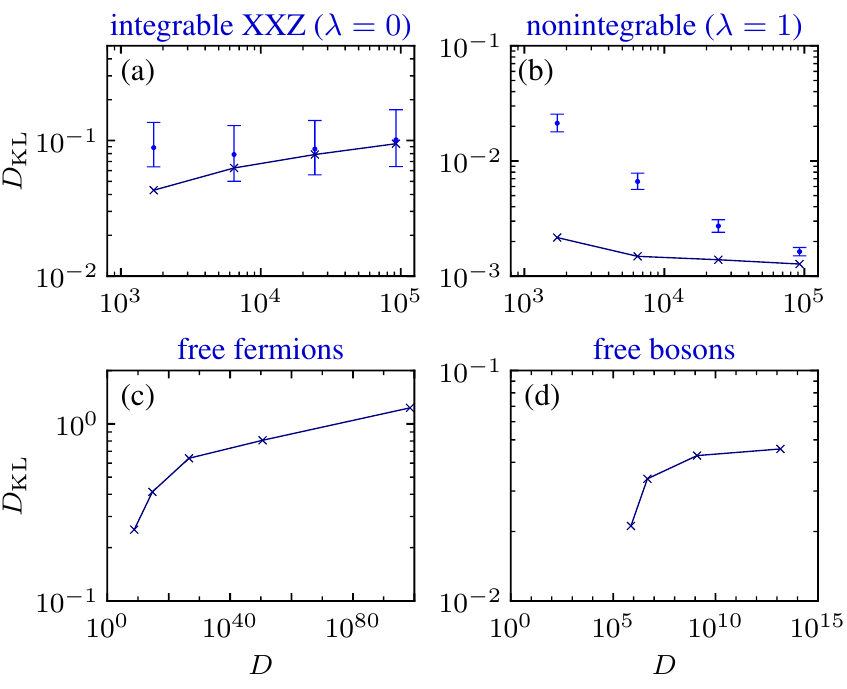}
  \caption{Kullback-Leibler divergence $D_\mathrm{KL}$ as function of the system size (Hilbert space
    dimension $D$).  (a,b) The XXZ chain ($\Delta=0.8$) for $250$ eigenstates in the middle of the
    spectrum.  The data points (dots) and error bars indicate the average and standard deviation of
    the per-state KLD. The solid line (crosses) shows the $D_\mathrm{KL}$ values of the distribution
    of the coefficients of the $250$ states taken together.  (c,d) Eigenstate-averaged KLD for free
    fermions and free bosons, respectively.  }
  \label{fig_xxz_sizes_dkl}
\end{figure}

Figure~\ref{fig_xxz_sizes_dkl} shows the KLD as a function of system size. The smallest and largest
system sizes for the XXZ chain [Figs.~\ref{fig_xxz_sizes_dkl}(a,b)] correspond to $L=13$ and $19$.
For the free-fermion and free-boson chains [Figs.~\ref{fig_xxz_sizes_dkl}(c,d)], the accessible
sizes are much larger, because the coefficient distributions can be obtained without explicit
numerical diagonalization of the many-body Hamiltonians, using the fact that each many-body
eigenstate is built out of single-particle eigenstates as a single Slater determinant
(non-interacting fermions) or as a single permanent (non-interacting bosons).  The issue is
discussed further in Sec.\ \ref{sec_integrable_numerics}.

In the non-integrable XXZ chain [Fig.~\ref{fig_xxz_sizes_dkl}(b)], there is a clear decrease of
$D_\mathrm{KL}$ (increasing Gaussianity) with increasing system size.
For the integrable cases, both the integrable XXZ chain and the non-interacting systems,
$D_\mathrm{KL}$ increases with system size, meaning that $P(z)$ becomes less Gaussian.  This is
consistent with our conjecture in the next section that $P(z)$ approaches a power law in the
large-size limit.  In addition, relatively large fluctuations between the eigenstates are observed
[error bars in Fig.~\ref{fig_xxz_sizes_dkl}(a)], reflecting the broad distribution seen in
Fig.~\ref{fig_manybody_coeff_dist}(b).  This is consistent with the idea that eigenstates of
integrable systems have a non-universal structure at finite sizes.

\section{Integrable many-body systems}
\label{sec_integrable}

We now concentrate on integrable systems and consider the coefficient distribution in the limit of
large sizes.  First, we describe the numerical analysis that leads to the conjecture of power-law
behavior at large sizes (Sec.\ \ref{sec_integrable_numerics}).  The rest of the section provides a series
of analytical arguments in support of this conjecture.

\subsection{Numerical analysis}
\label{sec_integrable_numerics}

\begin{figure*}
  \center%
  \includegraphics[width=170mm]{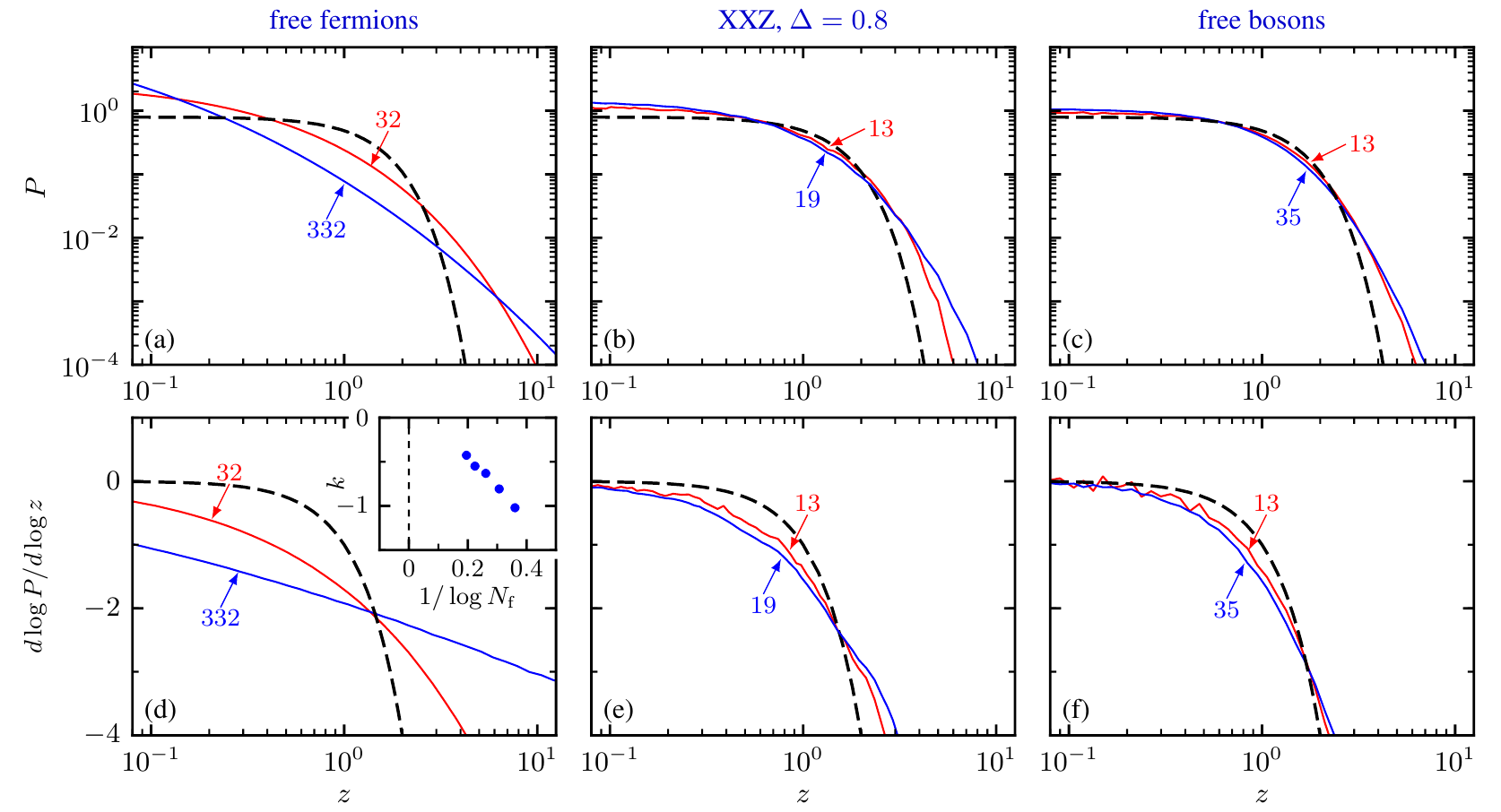}
  \caption{ (a--c) Amplitude distributions for integrable many-body systems, sampled from multiple
    eigenstates in the middle of the spectrum.  Dashed curves are Gaussians.  (a) Free-fermion
    chain; $(L,\nf)=(32,16)$ and $(332,166)$, ($D\sim 10^8$ and $10^{98}$).  (b) XXZ chain at
    $\Delta=0.8$, with $(L,\nup)=(13,6)$ and $(19,9)$ ($D=1716$ and $92378$).  (c) Free-boson chain;
    $(L,\nb)=(13,6)$ and $(35,17)$ ($D\sim10^4$ and $10^{13}$).  (d--f) Corresponding
    double-logarithmic derivatives.  Inset in (d) is the ``curvature'' $k=\frac{d^2\log P}{d(\log
      z)^2}$ as function of $1/\log \nf$ for the free-fermion chain.  }
  \label{fig_xxz_sizes}
\end{figure*}

The eigenstate coefficients for free bosons and free fermions can be evaluated
without explicit diagonalization of the many-body Hamiltonian, using the fact that the many-body
eigenstates are built out of single-particle eigenstates. The eigenstates are chosen by drawing the
`momenta' $k_j$ randomly such that the many-body energies lie in the desired energy range. 
For larger Hilbert spaces, we typically sample $10^3$--$10^4$ coefficients of each eigenstate.
For free fermions, the eigenfunctions are (Slater) determinants, which can be evaluated efficiently,
allowing us to sample relatively large systems ($>300$ sites).
For the XXZ chain, we are limited to exact diagonalization and the sizes are modest ($\approx20$
sites).  Intermediate are free bosons ($>30$ sites), whose eigenfunctions are permanents, whose
numerical evaluation is less favorable than determinants.

Figures~\ref{fig_xxz_sizes}(a,b,c) present double-logarithmic plots of the coefficient distribution
for three integrable systems.  A power-law behavior would show up as a straight line in this
representation.
The data in the free fermionic case shows a clear evolution toward power-law behavior as the system
size is increased.  The trend in the other two systems is in the same direction, but less
pronounced, presumably because of limited system sizes.  The data is further analyzed in
Figs.~\ref{fig_xxz_sizes}(d,e,f) through the slope of the curve in the double-logarithmic plot,
\begin{equation}\label{eqn_dld1def}
  \frac{d\log P}{d\log z} = \frac{z}{P}P'(z),
\end{equation}
the double-logarithmic derivative.  Power-law
behavior of $P$ would imply a constant (flat) double-logarithmic derivative, its value giving the power-law exponent.  The
inset shows that the slope of the double-logarithmic derivative in logarithmic scale,
\begin{equation}\label{eqn_dld2def}
 k=\frac{d^2\log P}{d(\log z)^2},
\end{equation}
becomes
smaller, arguably scaling to zero, in the large-size limit $\nf\to\infty$.  The available sizes do
not allow a meaningful extrapolation for the other two systems, but show the same general trend.
This observation, together with further supporting arguments below, lead to the conjecture that
the tails of the
coefficient distributions of eigenfunctions of integrable many-body systems approach power-law
shapes in the large-size limit.

\subsection{Structure of the many-body coefficients}

A common feature of several types of integrable systems is that many-body eigenstates can be
constructed out of single-particle eigenstates.  For example, a two-particle
wavefunction is of the form $\phi_a(1)\phi_b(2)\mp\phi_a(2)\phi_b(1)$ for fermions and bosons and of
the form $\phi_a(1)\phi_b(2)+e^{i\chi}\phi_a(2)\phi_b(1)$ for the XXZ chain, where $\chi$ is a phase
shift and $\phi_{a,b}$ are single-particle eigenstates.  We now consider the structure
of the many-body coefficients for larger system sizes, as to eventually obtain information
about their distribution.

Let us consider the single-particle Hamiltonian corresponding to the integrable many-body system of
interest.  This contains hopping terms and possibly a background potential,
\begin{equation}
H = \sum_{i=1}^{L-1}(a^\dagger_i a_{i+1}+a^\dagger_{i+1} a_{i})
  + \sum_{i=1}^L V_i a^\dagger_{i} a_{i}.
\end{equation}
(The geometry could be something other than a chain, e.g., a 2D or 3D lattice, and the hoppings
could be longer-range, without affecting any of the arguments below.)  The creation/annihilation
operators above can have any exchange statistics, for example, but not limited to, fermionic or
bosonic.

The creation operators for the single-particle eigenstates are linear combinations of the
$a^\dagger$ operators,
\begin{equation}\label{eqn_bose_eigenstates_op}
  d_k^\dagger = \sum_{j=1}^L \phi^{(k)}_j a_j^\dagger.
\end{equation}
Here, $k=1,\ldots,L$ labels the single-particle eigenstates, and the $j$ are site indices.
The $\phi^{(k)}_j$ are single-particle eigenstate coefficients.  For integrable systems, the
many-body eigenstate coefficients are built out of these  $\phi^{(k)}_j$'s.

In the simplest situation of nearest-neighbor hopping with no background potential, the
$\phi^{(k)}_j$ are sine functions, e.g., with open boundary conditions,
\begin{equation}
 \phi^{(k)}_j=\sqrt{\frac{2}{L+1}}\sin\frac{k j\pi}{L+1}.
\end{equation}
In this case, the indices $k$ can be
interpreted as momenta.  The corresponding single-particle energies are
$E^{(k)}=2\cos\frac{k\pi}{L+1}$.  The arguments below do not rely on a specific form of the
single-particle eigenstates and energies.

For non-interacting bosons or fermions, the many-body eigenstates are constructed by filling the
single-particle eigenstates with integer numbers of particles.  The eigenstates can be labeled
either as a list of occupancies of the $L$ single-particle eigenstates,
\begin{equation}
  \ket{\tilde{n}_1,\ldots,\tilde{n}_L} = \prod_{k=1}^L \frac{1}{\tilde{n}_k!} (d_k^\dagger)^{\tilde{n}_k}\ket{\text{vac}},
\end{equation}
or as a list of the single-particle eigenstates occupied by the $N$ particles,
\begin{equation}
  \ket{k_1,\ldots,k_\nn} = d_{k_1}^\dagger \cdots d_{k_\nn}^\dagger \ket{\text{vac}} .
\end{equation}
Here $\ket{\text{vac}}$ is the vacuum (no particles in the system). The integers $\tilde{n}_k\geq 0$
indicate how many particles are in single-particle eigenstate $\ket{k}$.  For fermions,
$\tilde{n}_k=0,1$ and for bosons they can take values up to $N$.  The many-body eigenenergy is equal
to $\sum_{k=1}^L \tilde{n}_k E^\mathrm{(k)}$.

For non-interacting bosons, the eigenstates can be expressed as the sum over permutations $p$ of the
positions $(j_1,\ldots,j_\nn)$ of the $\nn$ particles.
In the basis defined by the states $\ket{j_1,\ldots,j_\nn} \equiv a_{j_1}^\dagger \cdots a_{j_\nn}^\dagger\ket{\text{vac}}$, the eigenstate coefficients are
\begin{equation}\label{eqnCoeffFreeBosons}
  \braket{j_1,\ldots,j_\nn}{k_1,\ldots,k_\nn}\\
   = \frac{\sqrt{\gamma_{\{j\}}}}{\sqrt{\tilde{\gamma}_{\{k\}}}}
   \sum_{p\in\mathcal{P}}\phi^{(k_1)}_{p_1}\phi^{(k_2)}_{p_2}\cdots\phi^{(k_n)}_{p_n},
\end{equation}
where $\tilde{\gamma}_{\{k\}}=\prod_{k=1}^L {\tilde{n}_k!}$ and $\gamma_{\{j\}}=\prod_{j=1}^L {n_j!}$, and
$p=(p_1,\ldots,p_\nn)$ runs over all \emph{distinct} permutations of the particle positions
$(j_1,\ldots,j_\nn)$.  The summation may be conveniently implemented as the \emph{permanent} $\per
M$ of the $\nn\times \nn$ matrix $M$ defined by $M_{ab} = \phi^{(k_b)}_{j_a}$ ($a,b=1,\ldots,\nn$).

For free fermions, the many-body eigenstates are linear combinations of the products of the
single-particle eigenstates, as for free bosons.  However, antisymmetry under exchange of particles
introduces minus signs in this sum for odd permutations.  The coefficients are therefore given by
Slater determinants
\begin{align}\label{eqnCoeffFreeFermions}
  &\braket{j_1,\ldots,j_\nn}{k_1,\ldots,k_\nn}\\
  &\qquad
   = \frac{1}{\sqrt{\nn!}}\det M
   = \frac{1}{\sqrt{\nn!}} \sum_{p\in\mathcal{P}} (-1)^{p}
  \phi^{(k_1)}_{p_1}\phi^{(k_2)}_{p_2}\cdots\phi^{(k_n)}_{p_n}.\nonumber
\end{align}
The single-particle eigenstates $k_i$ are required to be distinct.

For systems solvable by the Bethe ansatz, the many-body amplitudes are also built out of
single-particle coefficients.  For the fermionic chain with nearest neighbor interactions
(equivalent to the XXZ chain for present purposes),
\begin{align}
  &\braket{j_1,\ldots,j_\nn}{k_1,\ldots,k_\nn}\nonumber\\
  &\qquad= \mathcal{N} \sum_{p\in\mathcal{P}}  \ee^{\ii\chi(p)} \;   \phi^{(k_1)}_{p_1}\phi^{(k_2)}_{p_2}\cdots\phi^{(k_n)}_{p_n},\label{eqnCoeffBetheAnsatz}
\end{align}
where the phase shift $\chi(p) = \sum_{i<j}\chi_2(k_i,k_j)$ is interaction dependent and is a sum of
two-particle phase shifts $\chi_2$, which are determined from the two-particle scattering problem.
For more complicated models, such as those requiring the nested Bethe ansatz, the wavefunction is
more involved, but the essential idea is the same.

\subsection{Argument for power-law behavior --- a toy model}
\label{sec_power_law_toy_model}

We provide an argument for power-law behavior of the coefficient distribution by considering
a toy model of $\nn$ \emph{distinguishable} particles, i.e., with
trivial exchange statistics. In this case the multi-particle eigenfunction coefficients 
\begin{equation}\label{eqn_product_coeff}
c =  \braket{j_1,\ldots,j_\nn}{k_1,\ldots,k_\nn}
  =\prod_{i=1}^\nn\braket{j_i}{k_i}
  =\prod_{i=1}^\nn\phi^{(k_i)}_{j_i} .
\end{equation}
are merely products of values of the single-particle eigenstates $\phi^{(k)}_{j}$, where $k$ label
the eigenstates and $j$ the site indices, cf.~\eqns\eqref{eqnCoeffFreeBosons}--\eqref{eqnCoeffBetheAnsatz}.
For $\nn$ distinguishable particles in $L$ sites, the Hilbert space dimension is $D = L^\nn$.

Because of the product-state nature of the coefficients $c$, it is natural to study the distributions
$Q(y)$ of the logarithms
\begin{equation}\label{eqn_defn_ylogz}
  y = \log{z} = \log{c} + \log\sqrt{D}.
\end{equation}
The distribution $Q(y)$ relates to the `usual' coefficient distribution $P(z)$ as $Q(y)=\ee^y P(\ee^y)$ and $P(z)=\frac{1}{z}Q(\log{z})$.  The
logarithm of the many-particle coefficients is the sum of the single-particle ones, $\log{c} =
\sum_{i=1}^\nn\log \phi^{(k_i)}_{j_i}$, so that
\begin{equation}\label{eqn_defn_mp_sp}
  y = \sum_{i=1}^\nn \log \phi^{(k_i)}_{j_i} +\tfrac{1}{2}\nn\log{L}.
\end{equation}
If we regard the single-particle coefficients to be effectively random, then this is a sum of $\nn$
random variables (plus a shift by a constant), and we can invoke the central limit theorem.  It
follows that $Q(y)$ at large $\nn$ approaches a Gaussian distribution,
\begin{equation}\label{eqnLogCoeffDist}
  Q(y) \to \frac{1}{\sqrt{2\pi\sigma_Q^2}}\ee^{-(y-\mu_Q)^2/2\sigma_Q^2},
\end{equation}
with mean $\mu_Q$ and variance $\sigma^2_Q$.  The central limit theorem yields the mean to be the
sum $\mu_Q = \sum_{i=1}^\nn \mu_{q_i}$ of the means $\mu_{q_i}$ of the single-particle
log-coefficient distributions $q_i$ of the variables $y_i=\log \phi^{(k_i)}_{j}+\frac{1}{2}\log L$
[i.e., one term in the sum of \eqn\eqref{eqn_defn_mp_sp}].  The term $\frac{1}{2}\log L$ represents
the scaling of the single-particle coefficients to unit variance, which renders the distributions
$q_i$ to be independent of system size in the limit $L\to\infty$.  The values $\mu_{q_i}$ only
depend on the lattice geometry and the quadratic couplings (e.g., short-range versus long-range
couplings).  Likewise, the variance $\sigma_Q^2$ approaches $\sum_{i=1}^\nn \sigma_{q_i}^2$ for
large $\nn$, where the $\sigma_{q_i}^2$ are the variances of the single-particle distributions
$q_i$.

The power-law behavior of $P(z)=\frac{1}{z}Q(\log{z})$ now readily follows from studying the first- and second-order double-logarithmic derivative,
\begin{align}
  &\frac{d \log P}{d\log z} =  -\frac{\log{z}}{\sigma_Q^2} + \frac{\mu_Q}{\sigma_Q^2} - 1,\label{eqnDLDAnalytic1}\\
  &k= \frac{d^2 \log P}{d(\log z)^2} = -\frac{1}{\sigma_Q^2}.\label{eqnDLDAnalytic2}
\end{align}
Under the assumption of identical distributions $q_i=q$, we have $\sigma_Q^2=\nn\sigma_q^2$ and
$\mu_Q=\nn\mu_q$ and find that the second-order double-logarithmic derivative scales as $k \sim 1/N$. Thus, for increasing system
size, the curvature of the coefficient distribution $P(z)$ on a double logarithmic scale decreases.
In other words, $P(z)$ `flattens' to a power law $\sim z^\alpha$. The exponent $\alpha$ of the
power law follows from the first-order double-logarithmic derivative, $\alpha\to\mu_q/\sigma_q^2 - 1$ (at $z=1$). The value
is non-universal: it is determined by the details of the single-particle coefficient distributions.

\begin{figure}
  \includegraphics[width=85mm]{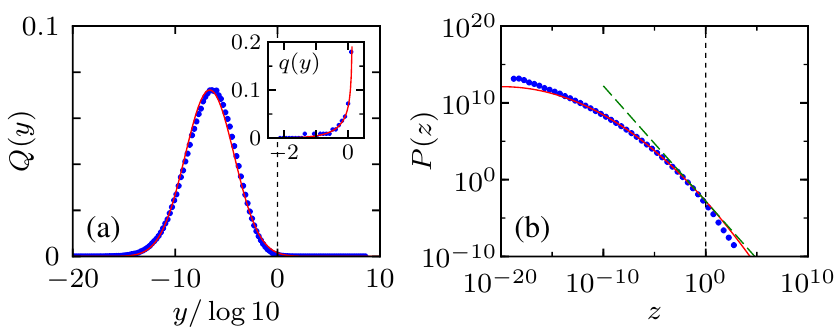}
  \caption{(a) Distribution $Q(y)$ of the logarithms of the coefficients of the
    many-particle-in-a-box model, for $(L,\nn)=(100,50)$. We consider a single eigenstate, with
    randomly chosen momenta $(k_1,\ldots,k_\nn)$. The (blue) dots indicate the numerical result
    obtained by sampling $10^6$ coefficients. The (red) curve is the estimated distribution $Q$ from
    \eqn\eqref{eqnLogCoeffDist}, with mean $\mu_Q$ and variance $\sigma_Q^2$ obtained from the
    single-particle distributions.  The inset shows a typical single-particle distribution $q(y)$,
    with the dots and curve indicating the discrete and continuum distributions, respectively.  (b)
    Corresponding coefficient distribution $P(z)$.  The (green) dashed line is an estimate for the
    power law with exponent evaluated at $z=1$.  }
  \label{figLogCoeffDist}
\end{figure}

In Fig.~\ref{figLogCoeffDist}, we illustrate the effectiveness of this argument with the results
for a model of many distinguishable particles in a finite chain with open boundary conditions
(many particles in a box). The numerically obtained
distribution fits well to the analytic estimate given by
\eqn\eqref{eqnLogCoeffDist} with $\mu_Q=\nn \mu_{q}$ and $\sigma_Q^2=\nn \sigma_{q}^2$.  (In this
case, the single-particle coefficient distributions are independent of the $k_i$; they are all
characterized by the same mean $\mu_{q}$ and variance $\sigma_{q}^2$.  We find $\mu_q\approx-0.347$
and $\sigma_q^2\approx 0.822$, which yields the exponent $\alpha\approx-1.42$.)  Small deviations
may be seen in the tails, and the numerical distribution is slightly skewed to the right, due to the
high asymmetry of the single-particle distribution $q$.  These deviations vanish in the limit
$\nn\to\infty$.

\subsection{Extension to non-trivial statistics}

From the data in Fig.~\ref{fig_xxz_sizes}, the question arises as of whether
the preceding argumentation for a large-$\nn$ approach to a power-law $P(z)$ can be extended
naturally to indistinguishable particles with non-trivial statistics.   For free bosons, free fermions
and systems integrable through the Bethe ansatz, the eigenfunctions are not just products of
single-particle wavefunctions, but linear combinations of them, as shown in
Eqs.\ \eqref{eqnCoeffFreeBosons}--\eqref{eqnCoeffBetheAnsatz},
respectively.  At present, we are able to outline a partial argument only for the
free-fermion case.

For free fermions, the many-body coefficients are determinants of the single-particle coefficients,
\eqn\eqref{eqnCoeffFreeFermions}.  Assuming these coefficients to be effectively random, we invoke
recent results from random matrix theory for the determinant of a random matrix
\cite{TaoVu2012,NguyenVu2014}.
Assuming that the entries of a matrix $A$ are essentially random, and their distribution is
sufficiently well-behaved, $\log\abs{\det A}$ satisfies a central-limit theorem: If the entries are
distributed with zero mean and unit variance, the distribution of $\log\abs{\det A}$ tends to a
normal distribution for large $\nn$, with mean $\frac{1}{2}\log (\nn-1)!$ and variance
$\frac{1}{2}\log \nn$.  For our matrix $M$ in \eqn\eqref{eqnCoeffFreeFermions}, the entries are
single-particle coefficients $\phi^{(k)}_j$ with variance $1/L$ by normalization.  The matrix
elements can be made to have unit variance by multiplying each element by $\sqrt{L}$, so that the
determinant is multiplied by $(\sqrt{L})^N$.  Thus, the coefficients are of the form
\begin{equation}
c =  \frac{1}{\sqrt{\nn!}} \frac{1}{L^{N/2}} \det{\tilde{M}},
\end{equation}
where the matrix $\tilde{M}$ now has entries with unit variance.  Unfortunately, the entries do not
necessarily have zero average; for example, if the coefficients are sinusoidal functions as in the
case of an open-boundary chain, half of the single-particle coefficients have nonzero average.  We
proceed with the assumption that this nonzero average causes a shift $\xi$ in the mean of the
distribution of  $\log\abs{\det \tilde{M}}$, and leaves the variance unchanged.  With this
assumption, the variable $y=\log\bigl(c\sqrt{D}\bigr)$ has a Gaussian distribution $Q(y)$, as in
\eqn\eqref{eqnLogCoeffDist}, with mean
\begin{align}
  \mu_Q &= - \log\sqrt{\nn!} - \log L^{N/2} \nonumber \\
  &\qquad+ \tfrac{1}{2}\log (\nn-1)!  + \xi(N) + \log\sqrt{D}
\end{align}
and variance $\sigma_Q^2=\frac{1}{2}\log \nn$.

Gaussianity of $Q(y)$ implies \eqns\eqref{eqnDLDAnalytic1} and \eqref{eqnDLDAnalytic2}
for the first- and second-order double-logarithmic derivative.
The latter
\begin{equation}
  k = -\frac{1}{\sigma_Q^2}  = \frac{1}{\frac{1}{2}\log\nn}
\end{equation}
thus vanishes at large $\nn$. This signifies an approach to power-law form
for $P(z)$ in the $\nn\to\infty$ limit.
Compared to the product-type states, the convergence is slower: $k\sim-1/\log \nf$
\cite{NguyenVu2014}.  This provides an appealing explanation to why we need enormous sizes to see
the approach to power-law behavior, and is the reason we plot $k$ against $1/\log\nf$ in
Fig.~\ref{fig_xxz_sizes}(d), inset.

We now attempt to estimate the power-law exponent at large sizes.  Equation~\eqref{eqnDLDAnalytic1}
implies that the exponent is
\begin{align}
  \frac{\mu_Q}{\sigma_Q^2}-1
  &= \frac{-\frac{1}{2}\log\nn - \log{L^{\nn/2}} + \log\sqrt{D} +
     \xi(\nn)}{\frac{1}{2}\log\nn} -1\nonumber\\
  &= -2 +  \frac{- \log{L^{\nn/2}} + \log\sqrt{D} + \xi(\nn)}{\frac{1}{2}\log\nn}.
\end{align}
The fraction (second term) is $N$-dependent.  (For the case of half-filling, $L=2\nn$, the numerator
is $-\frac{1}{2}\nn\log\nn+\xi(\nn)$ at large $\nn$.)  For a sensible large-$\nn$ limit, the $\nn$
dependence must be canceled by the unknown shift $\xi(\nn)$.  If the cancellation is perfect in the
sense that the fraction vanishes, we obtain the estimate $-2$ for the exponent, i.e., the asymptotic
power-law behavior $P(z)\propto z^{-2}$, which is consistent with the numerical data presented in
Fig.~\ref{fig_xxz_sizes}.  Of course, since we do not know the function $\xi(\nn)$,
the fraction could also be
an $\nn$-independent constant, in which case the exponent would be shifted from $-2$.

The assumption that the nonzero average of the matrix elements leads only to a shift in the mean of
$Q(y)$ seems quite reasonable.  Proving such an assumption, or deriving $\xi(\nn)$, is well beyond
the scope of the present work.  The central limit theorem for log-determinants, invoked above, is
cutting-edge mathematical work.  We are not aware of mathematical results with modified conditions
for the distribution of elements.  Note, however, that a power-law dependence can be inferred under
much weaker conditions than the assumption used above --- as long as $\sigma_Q^2$ is an increasing
function of $\nn$, we obtain a power-law $P(z)$ at large $\nn$.

At present, to our knowledge, no comparable central-limit-theorem analog is available for
permanents [\eqn\eqref{eqnCoeffFreeBosons}] and certainly not for more complicated generalizations
like \eqn\eqref{eqnCoeffBetheAnsatz} appearing in Bethe-ansatz wavefunctions, but a similar
Gaussian limit for $Q(y)$, and hence a power-law $P(z)$ for large $N$, seems plausible.
Thus, based on our numerical results and on the arguments above, a reasonable conjecture is that
$P(z)$ approaches a power law \emph{generically} in large-size integrable systems.

Any of these arguments (whether for trivial or for non-trivial statistics) rely on treating the
single-particle logarithmic coefficients $\log\phi_{a}$ as independent random variables.  While such
an assumption is likely impossible to be `proved', arguments in the same spirit underlie the
eigenstate thermalization hypothesis (ETH) and its extensions \cite{Deutsch1991, Srednicki1994,
  RigolEA2008, NeuenhahnMarquardt2012, BeugelingEA2014PRE, BeugelingEA2015PRE,
  BeugelingEA2015JStatMech, PolkovnikovRigol_AdvPhys2016, BorgonoviIzrailevSantos_PhysRep2016,
  MondainiRigol_PRE2017}.

\section{Quantum billiards}
\label{sec_billiards}

In this section, we consider single-particle systems (`billiards') confined to a two-dimensional
region either by a hard wall (Sec.\ \ref{subsec_billiard_hardwall}) or by a parabolic confining potential
(Sec.\ \ref{subsec_billiard_softwall}).  We show that a number of integrable billiards have amplitude
distributions with power-law tails, and present some data for systems with a mixed phase space.

\subsection{Hard walls}
\label{subsec_billiard_hardwall}

We now consider a single particle confined in a two-dimensional region $\Omega$.  The eigenfunctions
$\psi_n(\vec{x})$, $\vec{x} = (x,y)$, satisfy the Schr\"odinger equation
$-\nabla^2 \psi_n(\vec{x}) = E_n \psi_n(\vec{x})$
for $\vec{x}\in\Omega$, and vanish for $\vec{x}\not\in\Omega$.
Given an eigenstate $\psi(\vec{x})$ of a quantum billiard, we consider the probability distribution
$P(z)$ of the (rescaled) absolute values $z$ of the
amplitudes, $z = |\psi(\vec{x})|\sqrt{\mathcal{A}}$, i.e.,
\begin{equation}\label{eqn_def_amplitude_distribution}
  P(z)dz =  \frac{1}{\mathcal{A}}\, \int_{z\leq\left|\psi(x,y)\right|\sqrt{\mathcal{A}}<z+dz} 1\, dx dy.
\end{equation}
Here, $\mathcal{A}=\text{area}(\Omega)$ is the area allowed by the billiard potential.  Inclusion of
the factor $\sqrt{\mathcal{A}}$ ensures that $P(z)$ has unit variance.

\begin{figure}
  \center%
  \includegraphics[width=85mm]{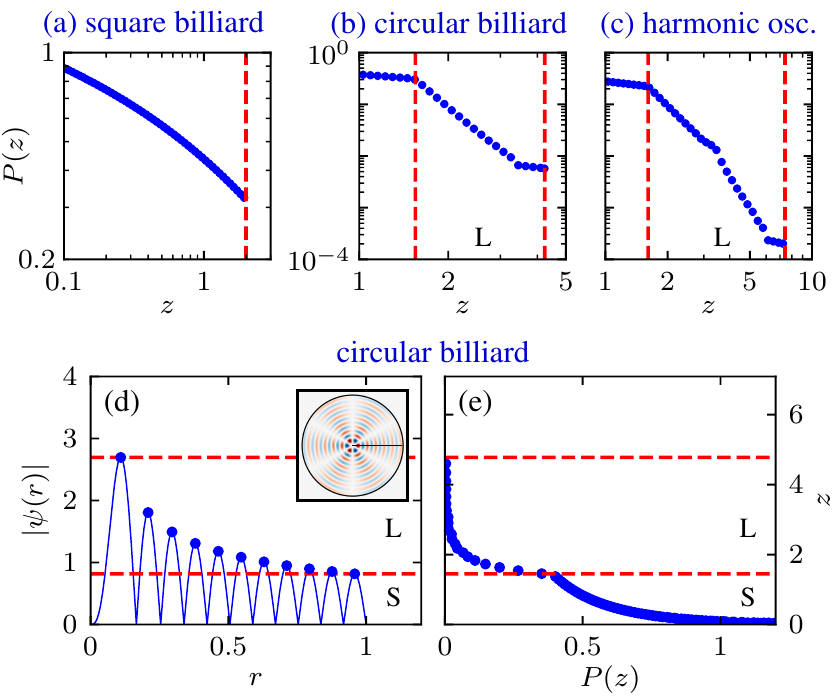}
  \caption{Amplitude distributions in single-particle systems (log-log plots). Distributions shown
    for a single eigenstate of: (a) The square billiard, $\psi^\square_{1,1}$.  (b) The circular
    billiard, $\psi^\circ_{101,37}$.  (c) An anisotropic two-dimensional harmonic oscillator,
    $\psi^\mathrm{ho}_{141,121}$.  (d,e) Illustration of the origin of two distinct regimes (labeled S and L) of the amplitude distribution; here, for the circular billiard, eigenfunction $\psi^\circ_{3,11}$. In (d), the curve indicates $\abs{\psi(r)}$ and the dots indicates local extrema.
    The inset visualizes $\psi^\circ_{3,11}(x,y)$.
    In (e), the resulting coefficient distribution is plotted sideways as function of $z$, which is scaled as to match the values of $\abs{\psi(r)}$ in (d).
    In all panels, the red dashed lines indicate the `L' regime, where power-law
    behavior can be expected. (For the square billiard, this regime is undefined.)  }
  \label{fig_ampdist}
\end{figure}

In contrast to chaotic quantum billiards, for which almost all eigenstates have Gaussian amplitude
distributions \cite{McDonaldKaufman_PRA1988, ShapiroGoelman_PRL1984, AurichSteiner_PhysicaD1991,
  AurichSteiner_PhysicaD1993, LiRobnik_JPA1994, Prosen_PhysLett97, Backer2007,
  SimmmelEckert_PhysicaD1996}, we here
investigate integrable billiards, as in \cite{SamajdarJain_integrablebilliards}. 
The simplest case is the square billiard, whose eigenfunctions $\psi^\square_{n_xn_y} \propto
\sin(n_x x) \sin(n_y y)$ all have the same amplitude distribution, which can be expressed analytically in terms of an elliptic integral (see Appendix~\ref{app_billiards}) and is shown in
Fig.~\ref{fig_ampdist}(a).
The tail of the distribution, which originates from the peaks of the
wavefunction, is $\sim z^{-1/2}$, but there is no extended power-law region.
The circular billiard eigenstates
$\psi^\circ_{mn}$ are labeled by angular and radial quantum numbers $m$ and $n$.  At large $n$, the
wavefunction has many oscillations in the radial direction, given by a Bessel function.  This leads
to a broad power-law segment in the amplitude distribution [see Fig.~\ref{fig_ampdist}(b)]: $P(z)\sim z^{-\gamma}$, with
$\gamma\approx5$ for $n\gg1$ (see Appendix~\ref{app_billiards}).
The region extends from the height of the lowest peak to
that of the highest peak, as illustrated by Figs.~\ref{fig_ampdist}(d,e).
In Fig.~\ref{fig_ampdist}, the expected power-law regime, defined by the minimum and maximum peak
amplitude, is indicated by the red dashed lines.

\subsection{Soft walls}
\label{subsec_billiard_softwall}

These results also extend to single-particle eigenstates of smooth confining potentials, i.e., with
Hamiltonian $H=-\nabla^2 + V(x,y)$.  In this case, we need to restrict the analysis of the
distribution to the classically accessible region, and define $\mathcal{A}$ [cf.~\eqn\eqref{eqn_def_amplitude_distribution}] to be the area of this region.
Let us consider the 2D harmonic oscillator, $V_{\textrm{h.o.}}(x,y)=x^2+\lambda^2y^2$.  The constant
$\lambda=\frac{1}{2}(1+\sqrt{5})$ is taken to be irrational in order to avoid complications with
degeneracies.  The eigenfunctions $\psi_{mn}^{\mathrm{h.o.}}$ are products of the eigenfunctions of
the one-dimensional harmonic oscillator, and have amplitude distributions with power-law tails. In
Fig.~\ref{fig_ampdist}(c), we illustrate the example $(m,n)=(141,121)$.
Like the circular billiard, this dependence arises
due to a combination of many inequivalent peaks in the wavefunctions.  The `kink' in the power law
regime at $z\approx 3$, that separates two regimes with different power law exponents,
presumably stems from the product structure.

\begin{figure}[tbp]
  \includegraphics[width=85mm]{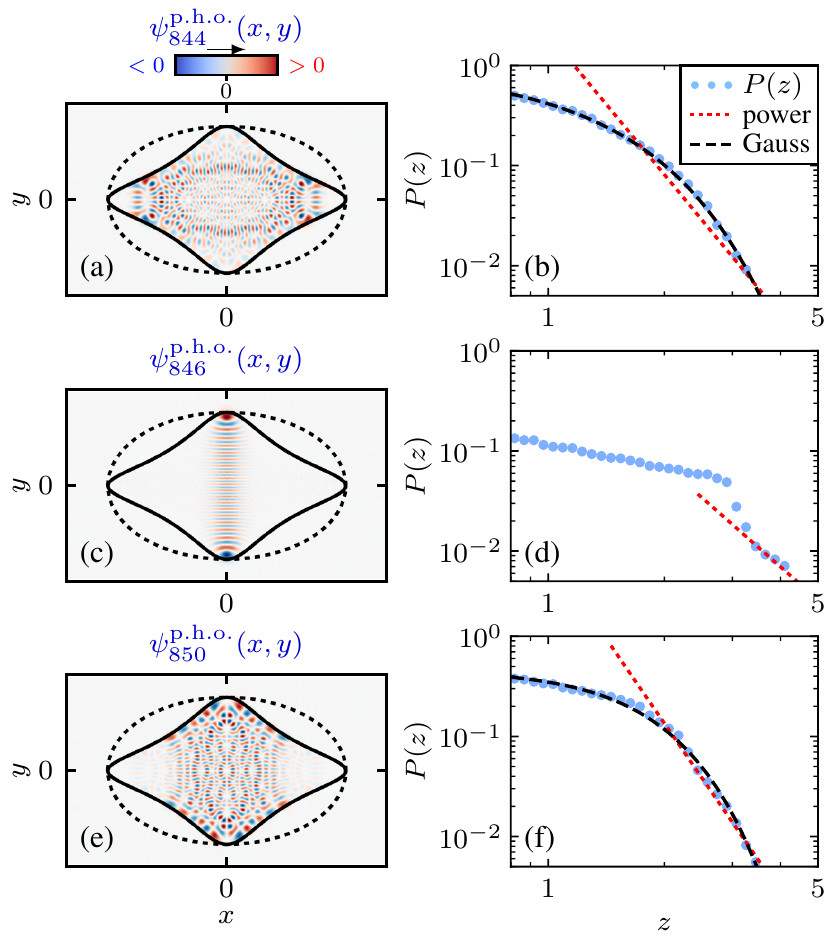}
  \caption{Eigenstates of perturbed anisotropic 2D harmonic oscillator.  The left-hand panels are
    wave functions $\psi^\mathrm{p.h.o.}_k(x,y)$ (labeled $k=0,1,2,\ldots$ in
    increasing order of eigenvalues). The right-hand panels are double logarithmic plots of
    amplitude distributions $P(z)$ (blue points). We also indicate Gaussian and power law fits.  In the wave-function plots, solid lines indicate the
    equipotential curve $V(x,y)=E^{(k)}$, and dotted elliptical lines indicate corresponding curves
    for the unperturbed potential.  (a,b) For the state $\psi^\mathrm{p.h.o.}_{844}$, $P(z)$ has a Gaussian
    tail.  (c,d) The state $\psi^\mathrm{p.h.o.}_{846}$, resembling a one-dimensional harmonic oscillator
    eigenstate, shows $P(z)$ features similar to that seen in integrable systems: a kink followed by
    an arguably power law tail.  (e,f) State $\psi^\mathrm{p.h.o.}_{850}$ is an intermediate state, whose
    tail fits neither a power law, nor a Gaussian, particularly well.}
  \label{figHarmOsc}
\end{figure}

Generically, if the potential is modified, many eigenfunctions at higher energies look chaotic and
have amplitude distributions with Gaussian tails.  Typically, the phase space at higher energies is
mainly chaotic, with small regular islands surrounding the short stable periodic orbits.  Thus, in
this regime, one typically encounters only a few regular eigenfunctions among many chaotic ones.  In order to
illustrate this observation, let us consider the weakly anharmonic potential 
$
  V(x,y)=x^2 +(\lambda y)^2 +\alpha x^2 y^2
$,
with $\alpha=0.2$.  In Fig.~\ref{figHarmOsc}(a,b), we show an example of a chaotic eigenstate, whose amplitude
distribution is Gaussian and whose eigenfunction is random-wave-like in a significant area of the 
classically accessible region.
At nearly equal energy, we also find an example of a highly regular eigenfunction
[Fig.~\ref{figHarmOsc}(c,d)];
typically, such eigenfunctions have a large overlap with a small number of eigenstates of the
non-perturbed model. The amplitude distribution typically shows
power-law tails after a kink, much like the $\ket{m,n}$ themselves.
Furthermore, there are
``intermediate'' eigenstates which have an extended wave function, but neither a power law nor a
Gaussian fits well [Fig~\ref{figHarmOsc}(e,f)].

\section {Context \& Conclusions}
\label{sec_conclusion}

We have extended the study of amplitude distributions to
many-body quantum systems.
One context for this work is a growing appreciation that concepts from the field of single-particle
quantum chaos can be useful for many-body quantum systems \cite{Deutsch1991, Srednicki1994,
  Horoi_Zelevinsky_PRL1995, ZelevinskyEA_PhysRep1996, Flambaum_Izrailev_Casati_PRE1996,
  Flambaum_Izrailev_PRE1997, Guhr_Weidemuller_PhysRep1998, RigolEA2008, RigolSantos2010,
  SantosRigol2010, SantosRigol2010a, SantosEA2012, PolkovnikovRigol_AdvPhys2016,
  BorgonoviIzrailevSantos_PhysRep2016}.
A global study of \emph{all} many-body eigenstates, such as the present one, is not historically
common in condensed matter physics.
The full eigenspectrum has gained importance only recently, due to intense interest in the dynamics
of isolated systems, including thermalization-related questions \cite{Deutsch1991, Srednicki1994,
  PolkovnikovRigol_AdvPhys2016, RigolEA2008, BorgonoviIzrailevSantos_PhysRep2016} and many-body
localization \cite{NandkishoreHuse_AnnuRev2015, AltmanVosk_AnnuRev2015}.

Our most striking result is the hint of a new type of universality associated with \emph{integrable}
many-body systems --- the coefficient distribution approaches a power-law in the large-size limit.
We have presented data and arguments to conjecture that this is a generic feature of multiple
classes of integrable systems.  Interestingly, we have shown that a number of regular
single-particle billiards also show power-law tails in $P(z)$, although we do not claim this to
be generic.  

For non-integrable many-body systems, away from the spectral edges, we have found Gaussian amplitude
distributions, as expected.  An interesting feature is the slight deviation from Gaussianity in 1D
geometries.
Gaussian behavior is a measure for the randomness of eigenstates and thus a characterization of
non-integrable behavior; in this respect it complements other eigenstate properties such as
entanglement randomness \cite{BeugelingEA2015JStatMech, VidmarRigol2017_entanglement}, inverse PR
\cite{BorgonoviIzrailevSantos_PhysRep2016, BeugelingEA2015JStatMech, SantosRigol2010,
  SantosRigol2010a}, phase correlators \cite{ArmstrongZelevinsky_PRE2012}, ETH scaling
\cite{BeugelingEA2014PRE, HaqueMcClarty_SYKETH} etc.  However, beyond its connection to chaos in
eigenstates, we regard the coefficient distribution to be an object of basic importance in its own
right.

This work raises a number of new questions:

(1) Ground states of many-body systems are multifractal \cite{AtasBogomolny_PRE12,
  LuitzAletLaflorencie_PRL14}.  Multifractality is related to the moments of $P(z)$; hence in light
of the present work one would like to investigate multifractality in the full spectrum.  This
requires modifying the definition in terms of size scaling, since there is no natural correspondence
between eigenstates of different-sized systems.  Appendix \ref{app_multifr} provides some data on
multifractality in a non-integrable many-body system.  

(2) The distributions for eigenstates near the spectral edge are clearly not Gaussian
(Fig.~\ref{fig_manybody_coeff_dist}), but it is unclear whether there is any generic behavior, or a
generic limiting distribution at large sizes.

(3) In (near-integrable) quantum billiards with a mixed phase space, the non-Gaussianity of $P(z)$ can sometimes be
described by, e.g., modified Gaussians with position dependent variance
\cite{BaeckerSchubert_JPA2002}.  For many-body systems, characterizing the non-Gaussianity of
near-integrable eigenstates remains an open task.

(4) The basis dependence of the distributions is an open question.  For example, in a mean-field
basis \cite{BorgonoviIzrailevSantos_PhysRep2016} such as the eigenbasis of the XX ($\Delta=0$)
Hamiltonian, the integrable XXZ chain has a high-kurtosis non-Gaussian amplitude distribution,
similar to the distribution in the coordinate basis [Fig.~\ref{fig_manybody_coeff_dist}(b)].

Each of these questions points to interesting directions for future study.


\appendix

\section{Comments on Quantum Integrability  \label{suppsec_integrability}}

In classical mechanics, the notion of integrability is commonly understood to mean the presence of
(at least) as many conserved quantities as the number of degrees of freedom
(``Liouville-integrability'').  In contrast, for quantum systems, there are a number of different
notions of integrability, and it is possible to find exceptions to or inadequacies with most
definitions.  We briefly discuss here a few notions associated with integrability, so that the sense
in which we have used the term is sufficiently clear.

Single-particle quantum billiard systems are called chaotic if the dynamics of the corresponding
classical Hamiltonian system is chaotic.  Conversely, a single-particle quantum system may be
regarded as integrable or regular if the corresponding classical system has integrable (`regular')
dynamics.  When the integrability is broken,
the system typically has a so-called mixed phase space, consisting of regions with regular
motion and regions with chaotic motion.
This is reflected in the quantum eigenstates which are
typically concentrated either within the regular regions or
the chaotic regions.

For quantum many-body systems, the situation is substantially more complicated
\cite{book_Sutherland_Beautiful_2004, CauxMossel_Integrability_JSTAT2011}.
Let us first consider systems where a large-size (`thermodynamic') limit is naturally defined, For
example, this includes fermionic or bosonic systems, where the limit is defined by increasing the
system size while keeping the density constant, and magnetic systems where the limit is defined by
increasing the system size while keeping the magnetization density constant.  In such cases, the
Hilbert-space dimension increases exponentially with system size.  A common notion of integrability
is that, if the system can be `solved' with polynomial rather than exponential effort, then the
system is integrable.  Here, `solving' means finding the eigenvalues and eigenstates of the
Hamiltonian.  For example, for systems of non-interacting fermions or bosons, it is sufficient to
find the eigenvalues and eigenstates of the single-particle problem; this allows construction of the
many-body eigenstates.  For systems solvable by the Bethe ansatz, the problem can be reduced to a
polynomial number of nonlinear equations.  In the simpler examples of the Bethe ansatz, such as the
XXZ chain, the number of equations is equal to the number of particles.  For more complicated cases,
such as those requiring a nested Bethe ansatz solution, the counting is more complicated, but the
basic idea of polynomial solvability still applies.

The idea of polynomial solvability is closely related to the physical idea that an integrable system
has a macroscopic number of conserved quantities.  The number of conserved quantities scales
polynomially (generally linearly) with the system size.  The conserved quantities for
non-interacting fermions/bosons are the occupancies of single-particle modes.  For systems
integrable only via the Bethe ansatz, the conserved quantities are often difficult to construct
explicitly, although their existence is guaranteed.

It is interesting to note that the above notion of integrability relies on large-size scaling, and
thus strictly speaking is not defined for a single fixed-size system, which is in sharp contrast to
the single-particle billiard case.  However, if a many-body Hamiltonian is integrable, then a `reasonably
large' system will show Poissonian level statistics.  While this is a phenomenological statement and
not very rigorous, it is sufficient for many purposes, and we can thus consider the level statistics
to provide an operational distinction between integrable and non-integrable many-body systems.  The
advantage of this viewpoint is that one can discuss integrability in both single-particle quantum
billiards and in many-body systems within the same framework.
Note, however, that even for the single-particle case there
are integrable systems not following Poissonian statistics;
these are usually considered as ``non-generic''.

There are various situations where the setup assumed here is not appropriate.  For example, there
are single-impurity problems which are solvable by Bethe ansatz, such as the Kondo model, the
Anderson impurity model, and the interacting resonant level model.  In these cases, a
constant-density scaling is not natural, as the impurity is localized in space.  In addition, these
models are generally integrable only for linearized bath dispersions, and the high-energy spectrum
of the linearized models may not be very physical.  Another unclear situation involves
zero-dimensional models, such as the two-site Bose-Hubbard model.  Although the solution of this
model can be written in Bethe ansatz form, the ansatz does not reduce an exponential problem,
because the Hilbert space has polynomial size to begin with (growing linearly with the particle
number).
In this work, we have ignored such special situations and restricted to many-body models where a clean and
natural thermodynamic limit can be defined.


\begin{figure}
  \center \includegraphics[width=86mm]{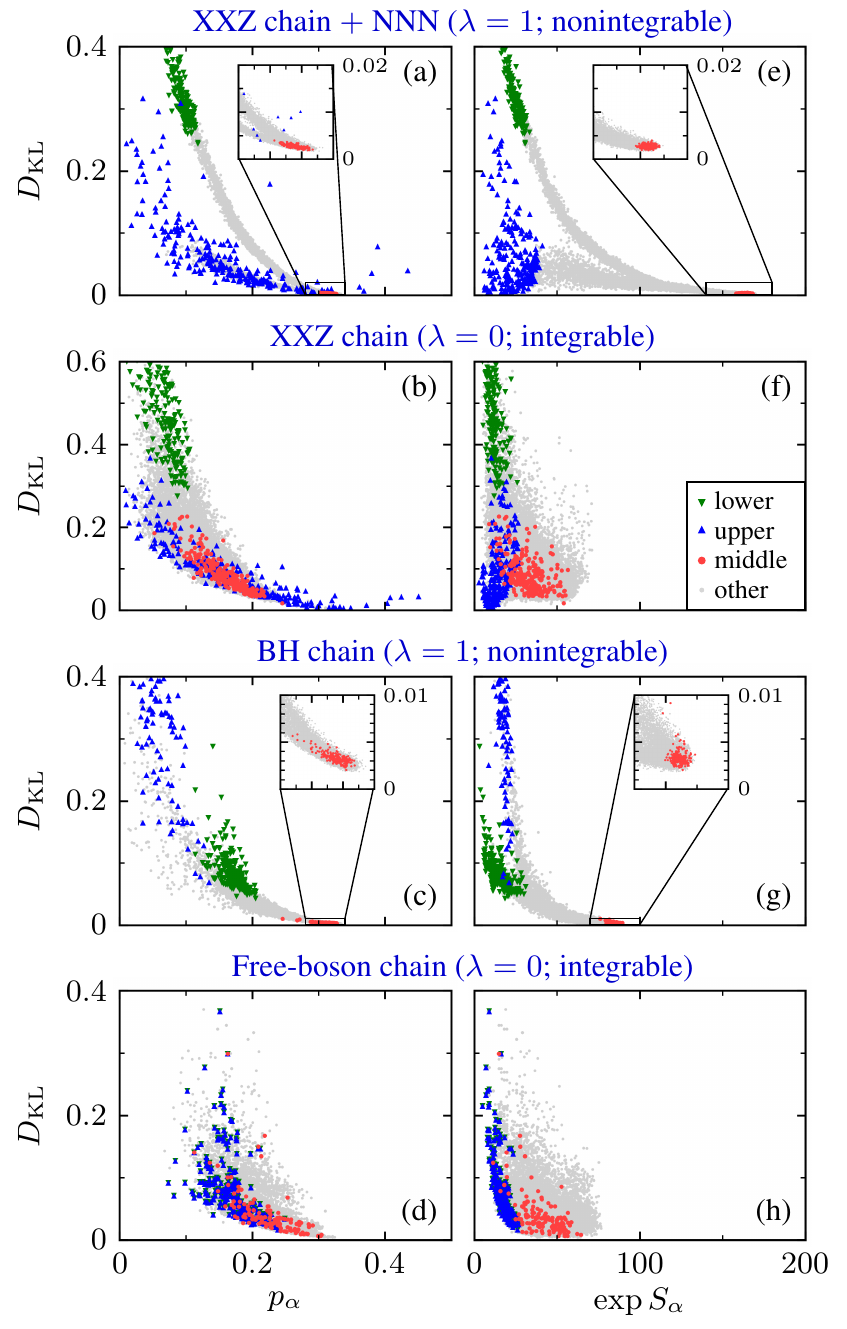}
  \caption{Kullback-Leibler divergence plotted against participation ratio (a)--(d) and against
    entanglement entropy with respect to left-right partition (e)--(h). We show data for the XXZ
    chain with and without NNN coupling, and for the boson chain with and without Hubbard
    interaction. The system sizes are $(L,\nup)=(17,8)$ and $(L,\nb)=(13,6)$, respectively. In all
    cases, we have highlighted $250$ states at the lower and upper edge of the spectrum and in the
    middle.}
  \label{fig_dkl_pr_ee}
\end{figure}

\section{Entanglement entropy, participation ratio, and deviations from Gaussianity}
\label{app_EE_PR}

In this Appendix we present data showing how the entanglement entropy (EE), the participation ratio
(PR) and the deviation from Gaussianity (quantified using the KLD) are correlated.

We first present scatter-plots (one data point for each eigenstate) allowing visualization of the
degree of correlation.  Figures~\ref{fig_dkl_pr_ee}(a)--(d) show plots of PR versus KLD for all
eigenstates, for the XXZ chain with and without NNN coupling, and for the bosonic model with and
without interaction.  Analogously, we show in Figs.~\ref{fig_dkl_pr_ee}(e)--(h) the correlation
between the KLD and $\exp(S_\mathrm{\alpha})$, where $S_\alpha$ is the EE with respect to a
partition of the system into two connected parts of sizes $l$ and $l+1$, where $2l+1=L$.  As argued
in Sec.~\ref{sec_many_body}, there is strong correlation visible in the non-integrable case [Figs.~\ref{fig_dkl_pr_ee}(a), (c), (e), and (g)]: a large
deviation from Gaussianity is associated with small PR and with small EE between spatial partitions.
The correlation is less clear in the integrable cases [Figs.~\ref{fig_dkl_pr_ee}(b), (d), (f), and (h)], similar to previous findings for the
correlation between EE and PR in eigenstates \cite{BeugelingEA2015JStatMech}.

\begin{figure}[tbp]
\center
\includegraphics[width=85mm]{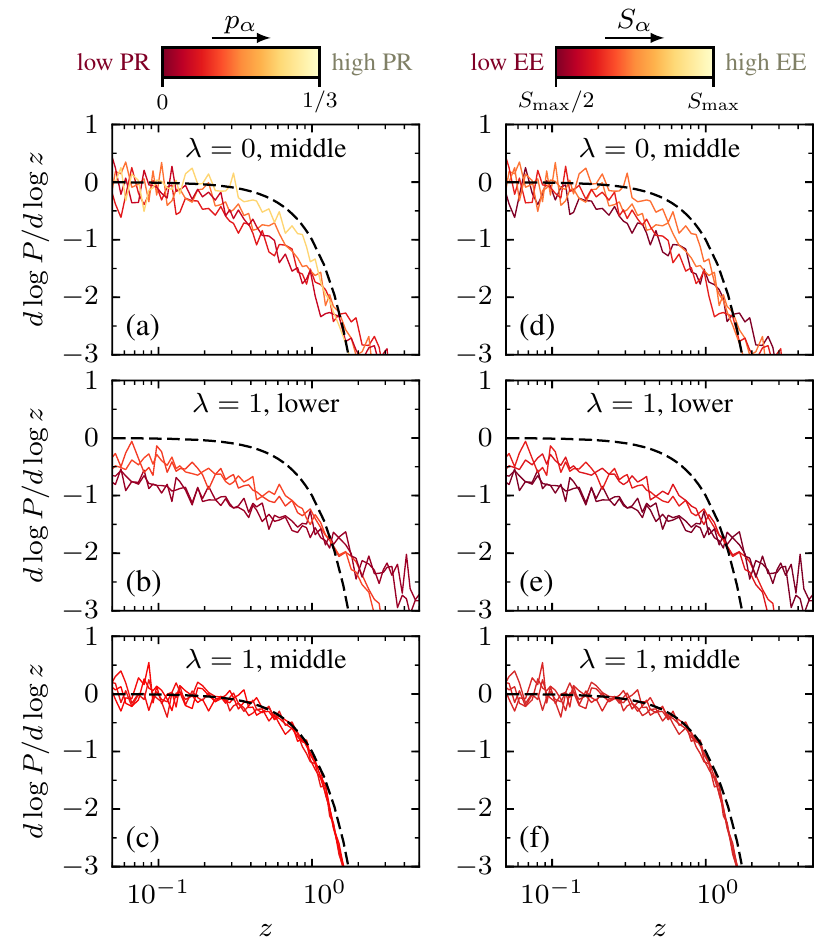}
\caption{Double-logarithmic derivative of the coefficient distribution for $4$ individual
  eigenstates.  (a,d) Integrable XXZ chain, middle of spectrum (near $E=0$). (b,e) Non-integrable
  XXZ chain, lower end of spectrum.  (c,f) Non-integrable XXZ chain, middle of spectrum.  Left
  column (a,b,c): Colors (shading) indicate values $P_\alpha$ of the PR.  The PR ranges from
  $P_\alpha=0$ to $P_\alpha=1/3$.  (d,e,f): The same distributions, but with the color (shading)
  representing EE values $S_\alpha$.  Values range from $S_\mathrm{max}/2$ to $S_\mathrm{max}$,
  where $S_\mathrm{max}=8\log 2$ is the maximum possible entropy. The dashed curves indicate the Gaussian distribution.
  }
  \label{fig_bystate}
\end{figure}

Next, in Fig.~\ref{fig_bystate}, we plot the double-logarithmic derivative of the coefficient
distributions for some of the individual eigenstates.  We have plotted the distributions of
$c^{(\alpha)}_\gamma$ of four individual representatives close to the designated part of the
spectrum. The chosen states are the highest-PR, lowest-PR, highest-EE, and lowest-EE eigenstates
within each group of $250$ eigenstates.

Comparing the curves, we observe a clear correlation between the shape of the distribution and the
PR and EE: The curves that lie closest to the Gaussian, e.g., for the non-integrable model in the
middle of the spectrum [Figs.~\ref{fig_bystate}(c) and (f); cf. Figs.~\ref{fig_dkl_pr_ee}(b) and
  (f)] are high-PR and high-EE states, as seen from the coloring (shading).  The other (flatter,
more power-law like) curves are low-PR and low-EE states.  For the non-integrable model, the
eigenstates close to the lower edge of the spectrum show integrable behavior.  This situation, shown
in Figs.~\ref{fig_bystate}(b) and (e) is very similar to generic eigenstates in the integrable
system [Figs.~\ref{fig_bystate}(a) and (d)].  On the other hand, in the bulk of the non-integrable
spectrum [Figs.~\ref{fig_bystate}(c) and (f)], all eigenstates have nearly Gaussian coefficient
distributions, and this is also reflected by the small variation in EE and PR.  With the KLD as a
measure of distance to Gaussian, these observations are consistent with those of
Figs.~\ref{fig_dkl_pr_ee}(a), (b), (e), and (f).


\section{Quantum billiards --- analytical observations}
\label{app_billiards}

In this Appendix, we provide some analytical results on amplitude distributions for the
single-particle (`quantum billiard') systems described in Sec.\ \ref{sec_billiards}, namely, the
square billiard and the circular billiard.

\subsection{Square billiard}

For the billiard in a square $\{(x, y) \mid 0\le x, y \le 1\}$ (so that $\mathcal{A}=1$)
with Dirichlet boundary conditions, the wave functions are given by
\begin{equation}
  \psisquare_{mn}(x,y) = 2\sin(m\pi x)\sin(n\pi y)
\end{equation}
with eigenenergies $\pi^2(m^2+n^2)$.  The coefficient distribution is independent of $m$ and $n$, so
we choose the ground state $m=n=1$ for simplicity and without loss of generality.

We present a derivation of $P(z)$ as the $z$ derivative of the area of the region defined by $\psi_{11}(x,y)<z$. We first simplify the
problem by studying the function $f(x,y)=\cos x \cos y$ (with $\abs{x},\abs{y}\leq \pi/2$), which is a scaled and shifted version of the wave function $\psi_{11}(x,y)$. The region defined by $f(x,y)>c$ encloses an area $A(c)$ complementary to the one we desire to compute. Considering the area in the first quadrant ($x\geq
0,y\geq0$) only, we find
\begin{equation}
  \tfrac{1}{4}A(c)=\int_0^{\arccos c}dy \arccos(c/\cos y),
\end{equation}
using that the boundary is given by $x=\arccos(c/\cos y)$, and $y$ runs from $0$ to $\arccos c$. Substitution $w=\cos y$, such that $dy=-dw/\sqrt{1-w^2}$, yields the integral
\begin{equation}\label{eqn_area_integral}
  \tfrac{1}{4}A(c)=\int_c^1 \frac{dw}{\sqrt{1-w^2}}\arccos (c/w).
\end{equation}
The solution to our initial problem, namely, the size of the level set defined by $z=\psi_{11}(x,y)$, is proportional to the derivative of $A(z/2)$, with a scaling factor $1/\pi^2$. By computation of the derivative of \eqn\eqref{eqn_area_integral}, we obtain
\begin{equation}\label{eqn_coeff_dist_analytic}
  P(z)=-\frac{1}{\pi^2}\frac{d}{dz}[A(z/2)]
      = \frac{2}{\pi^2}K\left(\sqrt{1-(z/2)^2}\right),
\end{equation}
where $K(k)=\int_0^1(1-t^2)^{-1/2}(1-k^2t^2)^{-1/2}dt$ denotes the complete elliptic integral of the first kind.

From this expression, we find
the approximate behavior $\sqrt{2/\pi^2 z}$ near the maximum value $z=2$, i.e., a scaling $\sim
z^{-\gamma}$ with $\gamma=\tfrac{1}{2}$.  There is however no extended power-law behavior.
Equation~\eqref{eqn_coeff_dist_analytic} also shows that the coefficient distribution diverges for
$z\to 0$.

\subsection{Circular billiard}

For the circular billiard of radius $1$ (with $\mathcal{A}=\pi$), the eigenstates of the Hamiltonian $H=-\nabla^2$ are given in polar coordinates $r,\phi$ by
\begin{equation}
  \psicircle_{mn}(\vec{r})=\mathcal{N}_{mn}J_{m}(j_{mn}r)\cos(m\phi)
\end{equation}
where $J_m$ is the Bessel function of the first kind of integer order $m\geq 0$, $j_{mn}$ is the
$n$'th zero of this function ($n>0$ integer),
and
\begin{equation}
  \mathcal{N}_{mn}^2
   =\left\{\begin{array}{ll}
     2/[\pi J_{m-1}(j_{mn})^2]\quad & (m>0)\\
     1/[\pi J_1(j_{0n})^2]\quad & (m=0)
    \end{array}\right.
\end{equation}
is the normalization factor.
The energy eigenvalue of this state is $j_{mn}^2$.

The amplitude distribution $P(z)$ has a power-law tail when the radial quantum number $n$ is large,
i.e., the eigenfunction has many oscillations in the radial direction.
The method of obtaining $P(z)$ can be illustrated using Figs.~\ref{fig_ampdist}(d,e).
The value $P(z)dz$ is
proportional to the area of the region where $\abs{\psi(\vec{r})}\in(c,c+dc)$,
with $c=z/\sqrt{\mathcal{A}}=z\pi^{-1/2}$, is satisfied.  For a
one-dimensional function $\psi(r)$, $P(z)$ is thus given by the sum over $1/\abs{\psi'(r_i)}$ over
all solutions $\psi(r_i)=c$.

As already addressed in Sec.~\ref{sec_billiards}, Figs.~\ref{fig_ampdist}(d,e) also visualize
two distinct regimes: For small $z$, $\abs{\psi(r,0)}=c=z\pi^{-1/2}$ has a fixed number of solutions,
and the coefficient distribution is
thus determined by the derivatives.  For larger $z$, the coefficient distribution is also affected
by the number of solutions, which gradually decreases if $z$ is increased.  The boundary between
these regimes is the value of the smallest local maximum of $\abs{\psi}$.

We will now focus on the limit of $m=0$ and large $n$. For large arguments $x$, the Bessel function
$J_m(x)$ behaves as an oscillatory function with amplitude $\sim x^{-1/2}$. More specifically (e.g.,
\eqn 9.2.1 of Ref. \cite{AbramowitzStegun1972}),
\begin{equation}\label{eqn_bessel_approx}
  J_m(x)=\sqrt{\frac{2}{\pi x}}\left[\cos(x-m\tfrac{\pi}{2}-\tfrac{\pi}{4})+\mathcal{O}(\abs{x}^{-1})\right].
\end{equation}
Given that $n$ is large, the argument $j_{0n}r$ in the Bessel function is large except for small
$r$.  The contribution from $r\to 0$ is suppressed due to the geometry (Jacobian of polar
coordinates); therefore it is reasonable to use the above large-$j_{0n}r$ approximation.  Thus
\begin{align}
  \psicircle_{0n}(\vec{r})
   &\approx\sqrt{\frac{1}{\pi J_1(j_{0n})^2}}\sqrt{\frac{2}{\pi j_{0n}r}}\cos (j_{0n}r-\tfrac{\pi}{4})\nonumber\\
   &\approx\sqrt{\frac{1}{\pi r}}\cos (n\pi r-\tfrac{\pi}{4}r-\tfrac{\pi}{4}),
   \label{eqn_wavefunction_approx}
\end{align}
where we have used $j_{0n}\approx (n-\frac{1}{4})\pi$ and $\pi J_1(j_{0n})^2\approx 2/j_{0n}$
according to approximation \eqref{eqn_bessel_approx}.

\emph{Small $z$---}
We first derive the behavior of $P(z)$ of $\psi^\circ_{0,n}$ for small $z$.
For $z=0$, we sum over all contributions where the wave function intersects zero. For this purpose,
we find the derivative
\begin{align}\label{eqn_psi0prime}
  \psi^{\circ\prime}_{0n}(r)
  &=-\tfrac{1}{2}\pi^{-1/2} r^{-3/2}\cos (n\pi r-\tfrac{\pi}{4}r-\tfrac{\pi}{4})\\
  &\quad{}-r^{-1/2}(n-\tfrac{1}{4})\pi^{1/2}\sin (n\pi r-\tfrac{\pi}{4}r-\tfrac{\pi}{4})\nonumber
\end{align}
with respect to $r$.
Let us label the zeros of $\psi_{0n}$ by $r_k=j_{0k}/j_{0n} \approx (k-\tfrac{1}{4})/(n-\tfrac{1}{4})$ with $k=1,\ldots,n$.
At $r=r_k$, the cosine term vanishes and the sine term is of unit magnitude, so that we find
\begin{align}
  \abs{\psi^{\circ\prime}_{0n}(r_k)}
  &=\sqrt{\pi/r_k}(n-\tfrac{1}{4})
  \approx \sqrt{\pi} (n-\tfrac{1}{4})^{3/2}(k-\tfrac{1}{4})^{-1/2}\nonumber\\
  &\approx \sqrt{\pi n^3/k}.
\end{align}
Summing over all zeros, the contribution to the coefficient distribution becomes (in approximation)
\begin{equation}\label{eqn_psi0_coeffdist_at0}
  2\sum_{k=1}^n 2\pi r_k / \abs{\psi'_{0n}(r_k)}
  \sim \sum_{k=1}^n \frac{k}{n} \frac{k^{1/2}}{n^{3/2}}
  = \sum_{k=1}^n \frac{k^{3/2}}{n^{5/2}} \sim 1.
\end{equation}
From this result, we deduce that the coefficient distribution has a finite value near $0$.  We find
the form $P(z)\approx\alpha +\beta z^2$ for small $z$ with $\alpha=\frac{8}{5}\sqrt{\pi}$ and
$\beta>0$.  The vanishing of the linear term in $z$ can be understood from a symmetry argument on
the coefficient distribution of $\psi_{0n}$, rather than of $\abs{\psi_{0n}}$.

\emph{Large $z$---}
When $z$ is larger than the smallest maximum, the number of intersections defined by
$\psi_{0n}(r)=c=z\pi^{-1/2}$ depends on the value of $z$.  The maxima of $\abs{\psi_{0n}(r)}$ are found at
$r'_k\approx (k+\tfrac{1}{4})/(n-\tfrac{1}{4})$ and are characterized by
$\abs{\psi_{0n}(r'_k)}\approx\sqrt{1/\pi r'_k} \approx\sqrt{n/k\pi}$. The number of intersections is
twice the number of maxima with $\abs{\psi_{0n}(r'_k)}\geq z$, i.e., $2 k_\mathrm{max}$ with
$k_\mathrm{max}=\lfloor n/\pi z^2\rfloor$.  At the intersection points the derivatives are also
roughly of the order $n^{3/2}k^{-1/2}$ (assuming that the sine term dominates, which is true for
$r\gg 1/n$, i.e., almost all $k$ except the smallest ones).  Then, performing a similar summation as
above, we obtain
\begin{align}
  2\sum_{k=1}^{k_\mathrm{max}} 2\pi r'_k / \abs{\psi'_{0n}(r'_k)}
  &\sim \sum_{k=1}^{k_\mathrm{max}} \frac{k^{3/2}}{n^{5/2}}
  \sim k_\mathrm{max}^{5/2}/n^{5/2}\nonumber\\
  &\sim (n/c^2)^{5/2}/n^{5/2}
  = c^{-5}
  \propto z^{-5}.
  \label{eqn_psi0_coeffdist_at2}
\end{align}
This scaling is valid for $z\gtrsim 1$ ($c\gtrsim\sqrt{1/\pi}\approx 0.56$), but the approximation becomes worse
for large $z$, i.e., if $z\sim\sqrt{n}$ ($c\sim \sqrt{n/\pi}$).  Numerical data (e.g., as shown in Fig.~\ref{fig_ampdist})
agrees with this finding: The coefficient distribution shows a power-law scaling $\alpha
z^{-\gamma}$ with $\gamma\approx 5$, with a deviation of less than $0.05$ for large $n$.  (For
example, $n=301$ yields $\gamma=-4.96\pm0.05$.)  The multiplicative constant $\alpha$ is almost
independent of $n$, because the coefficient distribution converges for $n\to\infty$: In this
limit the zeros and maxima of $\abs{\psi_{0n}}$ become denser, but the envelope remains unaltered,
cf.~\eqn\eqref{eqn_bessel_approx}.

For large $n$ and nonzero but small $m$, the numerical results show a similar scaling $z^{-\gamma}$,
where $\gamma$ is close to $5$.  The $\cos(m\phi)$ argument does not alter the overall derivation
outlined above.


\section{Multifractality \label{app_multifr}}

The moments of the coefficient distribution are used to define multifractality of wavefunctions.  It
is known that ground states of many-body systems are generally multifractal
\cite{AtasBogomolny_PRE12, LuitzAletLaflorencie_PRL14}.  We consider here extending this idea to the
full spectrum.
Following Ref.~\cite{AtasBogomolny_PRE12}, we define multifractality in terms of the R\'enyi
entropies obtained from the many-body wavefunction coefficients $c_\gamma$,
\begin{equation}
  S_\mathrm{R}(q,D) = -\frac{1}{q-1} \log\left(\sum_{\gamma=1}^D\abs{c_{\gamma}}^{2q}\right).
\end{equation}
The summation runs from $1$ to $D$, the Hilbert space dimension.
Then the fractal dimensions are defined as 
\begin{equation}
  \frdimq = \lim_{D\to\infty} \frac{S_\mathrm{R}(q,D)}{\log D}.
\end{equation}
Wavefunctions are \emph{multifractal} if the fractal dimension $\frdimq$ depends on the R\'enyi
parameter $q$.  They are simply \emph{fractal} if $\frdimq$ is a constant other than $1$.  For
Gaussian wavefunctions, they are expected to be constant at $\frdimq=1$.  Thus, we expect a
difference in scaling behavior between the spectral edges and the mid-spectrum eigenstates.
However, except for the ground state and the highest-energy state, it is not clear that there is a
meaningful comparison between sizes---an eigenstate for $L=13$ cannot in general be unambiguously
associated with an eigenstate of the $L=15$ system.  The $D\to\infty$ limit in the definition is
thus not \emph{a priori} well-defined.

\begin{figure}[bt]
\includegraphics[width=\columnwidth]{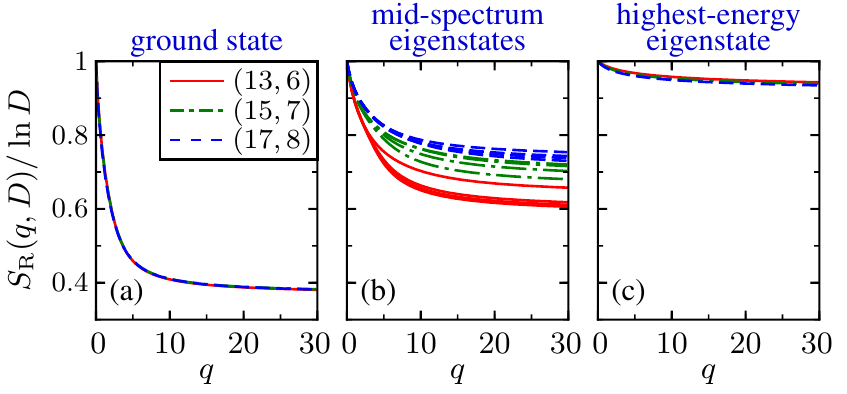}
\caption{Scaled R\'enyi entropies $S_\mathrm{R}(q,D)/\log{D}$ for (a) the ground state, (b) four
  mid-spectrum states, and (c) the highest-energy
  eigenstate of the XXZ model with NNN couplings.
  We compare three system sizes $(L,\nup)$, with $\nup=6,7,8$ up-spins in $L=2\nup+1$ sites; see the legend.
\label{fig_multifr01}}
\end{figure}

In Fig.~\ref{fig_multifr01} we display scaled R\'enyi entropies for the non-integrable spin chain
defined in \eqn\eqref{eqn_hxxz}, with $\lambda=1$ and $\Delta=0.8$.
We plot $S_\mathrm{R}(q,D)/\log{D}$, for the ground state, for four states near the middle of the
spectrum, and for the topmost (highest-energy) state.  In each case, three different system sizes
are compared.

The ground states and the topmost state of successive system sizes can be meaningfully compared, and
the limit in the definition of $\frdimq$ is unambiguous.  Both these cases
[Figs.~\ref{fig_multifr01}(a) and (c)] have almost converged already; extrapolation will give a
$q$-dependent fractal dimension.  Hence these non-Gaussian states are multifractal.  For the ground
state, this is consistent with the findings of Ref.~\cite{AtasBogomolny_PRE12}.

For other eigenstates, the limiting procedure is not well-defined, as explained above.  Here, we
simply take several eigenstates from near the center of the spectrum for each size,
Fig.~\ref{fig_multifr01}(b).  The trend as the size increases is consistent with the expected
non-fractal behavior ($\frdimq=1$).  However, the approach toward $\frdimq=1$ (assuming there is
such an approach) is quite slow.  Also, there are significant eigenstate-to-eigenstate fluctuations.
 
In summary, the data is consistent with the idea that the eigenstates are multifractal at the
spectral edges and non-fractal in the middle of the spectrum, but the limit is not unambiguously
defined and would be computationally challenging to characterize completely.
%

\acknowledgments

We thank A.~Chandran, M.~Heyl, I.~Khaymovich, A.~Lakshminarayan, C.~R.~Laumann, and L.~Santos for
discussions.


\begin{thebibliography}{65}%
\makeatletter
\providecommand \@ifxundefined [1]{%
 \@ifx{#1\undefined}
}%
\providecommand \@ifnum [1]{%
 \ifnum #1\expandafter \@firstoftwo
 \else \expandafter \@secondoftwo
 \fi
}%
\providecommand \@ifx [1]{%
 \ifx #1\expandafter \@firstoftwo
 \else \expandafter \@secondoftwo
 \fi
}%
\providecommand \natexlab [1]{#1}%
\providecommand \enquote  [1]{``#1''}%
\providecommand \bibnamefont  [1]{#1}%
\providecommand \bibfnamefont [1]{#1}%
\providecommand \citenamefont [1]{#1}%
\providecommand \href@noop [0]{\@secondoftwo}%
\providecommand \href [0]{\begingroup \@sanitize@url \@href}%
\providecommand \@href[1]{\@@startlink{#1}\@@href}%
\providecommand \@@href[1]{\endgroup#1\@@endlink}%
\providecommand \@sanitize@url [0]{\catcode `\\12\catcode `\$12\catcode
  `\&12\catcode `\#12\catcode `\^12\catcode `\_12\catcode `\%12\relax}%
\providecommand \@@startlink[1]{}%
\providecommand \@@endlink[0]{}%
\providecommand \url  [0]{\begingroup\@sanitize@url \@url }%
\providecommand \@url [1]{\endgroup\@href {#1}{\urlprefix }}%
\providecommand \urlprefix  [0]{URL }%
\providecommand \Eprint [0]{\href }%
\providecommand \doibase [0]{http://dx.doi.org/}%
\providecommand \selectlanguage [0]{\@gobble}%
\providecommand \bibinfo  [0]{\@secondoftwo}%
\providecommand \bibfield  [0]{\@secondoftwo}%
\providecommand \translation [1]{[#1]}%
\providecommand \BibitemOpen [0]{}%
\providecommand \bibitemStop [0]{}%
\providecommand \bibitemNoStop [0]{.\EOS\space}%
\providecommand \EOS [0]{\spacefactor3000\relax}%
\providecommand \BibitemShut  [1]{\csname bibitem#1\endcsname}%
\let\auto@bib@innerbib\@empty
\bibitem [{\citenamefont {Deutsch}(1991)}]{Deutsch1991}%
  \BibitemOpen
  \bibfield  {author} {\bibinfo {author} {\bibfnamefont {J.~M.}\ \bibnamefont
  {Deutsch}},\ }\href {\doibase 10.1103/PhysRevA.43.2046} {\bibfield  {journal}
  {\bibinfo  {journal} {Phys. Rev. A}\ }\textbf {\bibinfo {volume} {43}},\
  \bibinfo {pages} {2046} (\bibinfo {year} {1991})}\BibitemShut {NoStop}%
\bibitem [{\citenamefont {Srednicki}(1994)}]{Srednicki1994}%
  \BibitemOpen
  \bibfield  {author} {\bibinfo {author} {\bibfnamefont {M.}~\bibnamefont
  {Srednicki}},\ }\href {\doibase 10.1103/PhysRevE.50.888} {\bibfield
  {journal} {\bibinfo  {journal} {Phys. Rev. E}\ }\textbf {\bibinfo {volume}
  {50}},\ \bibinfo {pages} {888} (\bibinfo {year} {1994})}\BibitemShut
  {NoStop}%
\bibitem [{\citenamefont {Rigol}\ \emph {et~al.}(2008)\citenamefont {Rigol},
  \citenamefont {Dunjko},\ and\ \citenamefont {Olshanii}}]{RigolEA2008}%
  \BibitemOpen
  \bibfield  {author} {\bibinfo {author} {\bibfnamefont {M.}~\bibnamefont
  {Rigol}}, \bibinfo {author} {\bibfnamefont {V.}~\bibnamefont {Dunjko}}, \
  and\ \bibinfo {author} {\bibfnamefont {M.}~\bibnamefont {Olshanii}},\ }\href
  {\doibase 10.1038/nature06838} {\bibfield  {journal} {\bibinfo  {journal}
  {Nature}\ }\textbf {\bibinfo {volume} {452}},\ \bibinfo {pages} {854}
  (\bibinfo {year} {2008})}\BibitemShut {NoStop}%
\bibitem [{\citenamefont {D'Alessio}\ \emph {et~al.}(2016)\citenamefont
  {D'Alessio}, \citenamefont {Kafri}, \citenamefont {Polkovnikov},\ and\
  \citenamefont {Rigol}}]{PolkovnikovRigol_AdvPhys2016}%
  \BibitemOpen
  \bibfield  {author} {\bibinfo {author} {\bibfnamefont {L.}~\bibnamefont
  {D'Alessio}}, \bibinfo {author} {\bibfnamefont {Y.}~\bibnamefont {Kafri}},
  \bibinfo {author} {\bibfnamefont {A.}~\bibnamefont {Polkovnikov}}, \ and\
  \bibinfo {author} {\bibfnamefont {M.}~\bibnamefont {Rigol}},\ }\href
  {\doibase 10.1080/00018732.2016.1198134} {\bibfield  {journal} {\bibinfo
  {journal} {Adv. Phys.}\ }\textbf {\bibinfo {volume} {65}},\ \bibinfo {pages}
  {239} (\bibinfo {year} {2016})}\BibitemShut {NoStop}%
\bibitem [{\citenamefont {{Borgonovi}}\ \emph {et~al.}(2016)\citenamefont
  {{Borgonovi}}, \citenamefont {{Izrailev}}, \citenamefont {{Santos}},\ and\
  \citenamefont {{Zelevinsky}}}]{BorgonoviIzrailevSantos_PhysRep2016}%
  \BibitemOpen
  \bibfield  {author} {\bibinfo {author} {\bibfnamefont {F.}~\bibnamefont
  {{Borgonovi}}}, \bibinfo {author} {\bibfnamefont {F.~M.}\ \bibnamefont
  {{Izrailev}}}, \bibinfo {author} {\bibfnamefont {L.~F.}\ \bibnamefont
  {{Santos}}}, \ and\ \bibinfo {author} {\bibfnamefont {V.~G.}\ \bibnamefont
  {{Zelevinsky}}},\ }\href {\doibase 10.1016/j.physrep.2016.02.005} {\bibfield
  {journal} {\bibinfo  {journal} {Phys. Rep.}\ }\textbf {\bibinfo {volume}
  {626}},\ \bibinfo {pages} {1} (\bibinfo {year} {2016})}\BibitemShut {NoStop}%
\bibitem [{\citenamefont {Shapiro}\ and\ \citenamefont
  {Goelman}(1984)}]{ShapiroGoelman_PRL1984}%
  \BibitemOpen
  \bibfield  {author} {\bibinfo {author} {\bibfnamefont {M.}~\bibnamefont
  {Shapiro}}\ and\ \bibinfo {author} {\bibfnamefont {G.}~\bibnamefont
  {Goelman}},\ }\href {\doibase 10.1103/PhysRevLett.53.1714} {\bibfield
  {journal} {\bibinfo  {journal} {Phys. Rev. Lett.}\ }\textbf {\bibinfo
  {volume} {53}},\ \bibinfo {pages} {1714} (\bibinfo {year}
  {1984})}\BibitemShut {NoStop}%
\bibitem [{\citenamefont {McDonald}\ and\ \citenamefont
  {Kaufman}(1988)}]{McDonaldKaufman_PRA1988}%
  \BibitemOpen
  \bibfield  {author} {\bibinfo {author} {\bibfnamefont {S.~W.}\ \bibnamefont
  {McDonald}}\ and\ \bibinfo {author} {\bibfnamefont {A.~N.}\ \bibnamefont
  {Kaufman}},\ }\href {\doibase 10.1103/PhysRevA.37.3067} {\bibfield  {journal}
  {\bibinfo  {journal} {Phys. Rev. A}\ }\textbf {\bibinfo {volume} {37}},\
  \bibinfo {pages} {3067} (\bibinfo {year} {1988})}\BibitemShut {NoStop}%
\bibitem [{\citenamefont {{Aurich}}\ and\ \citenamefont
  {{Steiner}}(1991)}]{AurichSteiner_PhysicaD1991}%
  \BibitemOpen
  \bibfield  {author} {\bibinfo {author} {\bibfnamefont {R.}~\bibnamefont
  {{Aurich}}}\ and\ \bibinfo {author} {\bibfnamefont {F.}~\bibnamefont
  {{Steiner}}},\ }\href {\doibase 10.1016/0167-2789(91)90097-S} {\bibfield
  {journal} {\bibinfo  {journal} {Physica D}\ }\textbf {\bibinfo {volume}
  {48}},\ \bibinfo {pages} {445} (\bibinfo {year} {1991})}\BibitemShut
  {NoStop}%
\bibitem [{\citenamefont {Aurich}\ and\ \citenamefont
  {Steiner}(1993)}]{AurichSteiner_PhysicaD1993}%
  \BibitemOpen
  \bibfield  {author} {\bibinfo {author} {\bibfnamefont {R.}~\bibnamefont
  {Aurich}}\ and\ \bibinfo {author} {\bibfnamefont {F.}~\bibnamefont
  {Steiner}},\ }\href@noop {} {\bibfield  {journal} {\bibinfo  {journal}
  {Physica D}\ }\textbf {\bibinfo {volume} {64}},\ \bibinfo {pages} {185}
  (\bibinfo {year} {1993})}\BibitemShut {NoStop}%
\bibitem [{\citenamefont {{Li}}\ and\ \citenamefont
  {{Robnik}}(1994)}]{LiRobnik_JPA1994}%
  \BibitemOpen
  \bibfield  {author} {\bibinfo {author} {\bibfnamefont {B.}~\bibnamefont
  {{Li}}}\ and\ \bibinfo {author} {\bibfnamefont {M.}~\bibnamefont
  {{Robnik}}},\ }\href {\doibase 10.1088/0305-4470/27/16/017} {\bibfield
  {journal} {\bibinfo  {journal} {J. Phys. A: Math. Gen.}\ }\textbf {\bibinfo
  {volume} {27}},\ \bibinfo {pages} {5509} (\bibinfo {year}
  {1994})}\BibitemShut {NoStop}%
\bibitem [{\citenamefont {Simmel}\ and\ \citenamefont
  {Eckert}(1996)}]{SimmmelEckert_PhysicaD1996}%
  \BibitemOpen
  \bibfield  {author} {\bibinfo {author} {\bibfnamefont {F.}~\bibnamefont
  {Simmel}}\ and\ \bibinfo {author} {\bibfnamefont {M.}~\bibnamefont
  {Eckert}},\ }\href {\doibase 10.1016/0167-2789(96)00040-1} {\bibfield
  {journal} {\bibinfo  {journal} {Physica D}\ }\textbf {\bibinfo {volume}
  {97}},\ \bibinfo {pages} {517} (\bibinfo {year} {1996})}\BibitemShut
  {NoStop}%
\bibitem [{\citenamefont {Prosen}(1997)}]{Prosen_PhysLett97}%
  \BibitemOpen
  \bibfield  {author} {\bibinfo {author} {\bibfnamefont {T.}~\bibnamefont
  {Prosen}},\ }\href {\doibase http://dx.doi.org/10.1016/S0375-9601(97)00492-1}
  {\bibfield  {journal} {\bibinfo  {journal} {Phys. Lett. A}\ }\textbf
  {\bibinfo {volume} {233}},\ \bibinfo {pages} {332} (\bibinfo {year}
  {1997})}\BibitemShut {NoStop}%
\bibitem [{\citenamefont {{B{\"a}cker}}(2003)}]{Backer2003}%
  \BibitemOpen
  \bibfield  {author} {\bibinfo {author} {\bibfnamefont {A.}~\bibnamefont
  {{B{\"a}cker}}},\ }in\ \href@noop {} {\emph {\bibinfo {booktitle} {The
  Mathematical Aspects of Quantum Maps}}},\ \bibinfo {series} {Lecture Notes in
  Physics, Berlin Springer Verlag}, Vol.\ \bibinfo {volume} {618},\ \bibinfo
  {editor} {edited by\ \bibinfo {editor} {\bibfnamefont {M.~D.}\ \bibnamefont
  {{Esposti}}}\ and\ \bibinfo {editor} {\bibfnamefont {S.}~\bibnamefont
  {{Graffi}}}}\ (\bibinfo {year} {2003})\ pp.\ \bibinfo {pages} {91--144},\
  \Eprint {http://arxiv.org/abs/arXiv:nlin/0204061} {arXiv:nlin/0204061}
  \BibitemShut {NoStop}%
\bibitem [{\citenamefont {B{\"a}cker}(2007)}]{Backer2007}%
  \BibitemOpen
  \bibfield  {author} {\bibinfo {author} {\bibfnamefont {A.}~\bibnamefont
  {B{\"a}cker}},\ }\href {\doibase 10.1140/epjst/e2007-00153-4} {\bibfield
  {journal} {\bibinfo  {journal} {Eur. Phys. J. Special Topics}\ }\textbf
  {\bibinfo {volume} {145}},\ \bibinfo {pages} {161} (\bibinfo {year}
  {2007})}\BibitemShut {NoStop}%
\bibitem [{\citenamefont {Sutherland}(2004)}]{book_Sutherland_Beautiful_2004}%
  \BibitemOpen
  \bibfield  {author} {\bibinfo {author} {\bibfnamefont {B.}~\bibnamefont
  {Sutherland}},\ }\href {https://books.google.ie/books?id=PA0xV0XxZIwC} {\emph
  {\bibinfo {title} {Beautiful Models: 70 Years of Exactly Solved Quantum
  Many-body Problems}}}\ (\bibinfo  {publisher} {World Scientific},\ \bibinfo
  {year} {2004})\BibitemShut {NoStop}%
\bibitem [{\citenamefont {Caux}\ and\ \citenamefont
  {Mossel}()}]{CauxMossel_Integrability_JSTAT2011}%
  \BibitemOpen
  \bibfield  {author} {\bibinfo {author} {\bibfnamefont {J.-S.}\ \bibnamefont
  {Caux}}\ and\ \bibinfo {author} {\bibfnamefont {J.}~\bibnamefont {Mossel}},\
  }\href {\doibase 10.1088/1742-5468/2011/02/P02023} {\bibfield  {journal}
  {\bibinfo  {journal} {J. Stat. Mech.}\ }\textbf {\bibinfo {volume} {2011}},\
  \bibinfo {pages} {P02023}}\BibitemShut {NoStop}%
\bibitem [{\citenamefont {Bohigas}\ \emph {et~al.}(1984)\citenamefont
  {Bohigas}, \citenamefont {Giannoni},\ and\ \citenamefont
  {Schmit}}]{BohigasEA_PRL1984}%
  \BibitemOpen
  \bibfield  {author} {\bibinfo {author} {\bibfnamefont {O.}~\bibnamefont
  {Bohigas}}, \bibinfo {author} {\bibfnamefont {M.~J.}\ \bibnamefont
  {Giannoni}}, \ and\ \bibinfo {author} {\bibfnamefont {C.}~\bibnamefont
  {Schmit}},\ }\href {\doibase 10.1103/PhysRevLett.52.1} {\bibfield  {journal}
  {\bibinfo  {journal} {Phys. Rev. Lett.}\ }\textbf {\bibinfo {volume} {52}},\
  \bibinfo {pages} {1} (\bibinfo {year} {1984})}\BibitemShut {NoStop}%
\bibitem [{\citenamefont {Berry}\ and\ \citenamefont
  {Tabor}(1977)}]{BerryTabor_ProcRoy1977}%
  \BibitemOpen
  \bibfield  {author} {\bibinfo {author} {\bibfnamefont {M.~V.}\ \bibnamefont
  {Berry}}\ and\ \bibinfo {author} {\bibfnamefont {M.}~\bibnamefont {Tabor}},\
  }\href {\doibase 10.1098/rspa.1977.0140} {\bibfield  {journal} {\bibinfo
  {journal} {Proc. R. Soc. A}\ }\textbf {\bibinfo {volume} {356}},\ \bibinfo
  {pages} {375} (\bibinfo {year} {1977})}\BibitemShut {NoStop}%
\bibitem [{\citenamefont {Montambaux}\ \emph {et~al.}(1993)\citenamefont
  {Montambaux}, \citenamefont {Poilblanc}, \citenamefont {Bellissard},\ and\
  \citenamefont {Sire}}]{Montambaux_Poilblanc_PRL1993}%
  \BibitemOpen
  \bibfield  {author} {\bibinfo {author} {\bibfnamefont {G.}~\bibnamefont
  {Montambaux}}, \bibinfo {author} {\bibfnamefont {D.}~\bibnamefont
  {Poilblanc}}, \bibinfo {author} {\bibfnamefont {J.}~\bibnamefont
  {Bellissard}}, \ and\ \bibinfo {author} {\bibfnamefont {C.}~\bibnamefont
  {Sire}},\ }\href {\doibase 10.1103/PhysRevLett.70.497} {\bibfield  {journal}
  {\bibinfo  {journal} {Phys. Rev. Lett.}\ }\textbf {\bibinfo {volume} {70}},\
  \bibinfo {pages} {497} (\bibinfo {year} {1993})}\BibitemShut {NoStop}%
\bibitem [{\citenamefont {Poilblanc}\ \emph {et~al.}(1993)\citenamefont
  {Poilblanc}, \citenamefont {Ziman}, \citenamefont {Bellissard}, \citenamefont
  {Mila},\ and\ \citenamefont
  {Montambaux}}]{Poilblanc_Ziman_Mila_Montambaux_EPL1993}%
  \BibitemOpen
  \bibfield  {author} {\bibinfo {author} {\bibfnamefont {D.}~\bibnamefont
  {Poilblanc}}, \bibinfo {author} {\bibfnamefont {T.}~\bibnamefont {Ziman}},
  \bibinfo {author} {\bibfnamefont {J.}~\bibnamefont {Bellissard}}, \bibinfo
  {author} {\bibfnamefont {F.}~\bibnamefont {Mila}}, \ and\ \bibinfo {author}
  {\bibfnamefont {G.}~\bibnamefont {Montambaux}},\ }\href
  {http://stacks.iop.org/0295-5075/22/i=7/a=010} {\bibfield  {journal}
  {\bibinfo  {journal} {Europhys. Lett.}\ }\textbf {\bibinfo {volume} {22}},\
  \bibinfo {pages} {537} (\bibinfo {year} {1993})}\BibitemShut {NoStop}%
\bibitem [{\citenamefont {Hsu}\ and\ \citenamefont
  {Angl{\`e}s~d'Auriac}(1993)}]{HsuEA_PRB1993}%
  \BibitemOpen
  \bibfield  {author} {\bibinfo {author} {\bibfnamefont {T.~C.}\ \bibnamefont
  {Hsu}}\ and\ \bibinfo {author} {\bibfnamefont {J.~C.}\ \bibnamefont
  {Angl{\`e}s~d'Auriac}},\ }\href {\doibase 10.1103/PhysRevB.47.14291}
  {\bibfield  {journal} {\bibinfo  {journal} {Phys. Rev. B}\ }\textbf {\bibinfo
  {volume} {47}},\ \bibinfo {pages} {14291} (\bibinfo {year}
  {1993})}\BibitemShut {NoStop}%
\bibitem [{\citenamefont {Rabson}\ \emph {et~al.}(2004)\citenamefont {Rabson},
  \citenamefont {Narozhny},\ and\ \citenamefont
  {Millis}}]{NarozhnyMillis_PRB2004}%
  \BibitemOpen
  \bibfield  {author} {\bibinfo {author} {\bibfnamefont {D.~A.}\ \bibnamefont
  {Rabson}}, \bibinfo {author} {\bibfnamefont {B.~N.}\ \bibnamefont
  {Narozhny}}, \ and\ \bibinfo {author} {\bibfnamefont {A.~J.}\ \bibnamefont
  {Millis}},\ }\href {\doibase 10.1103/PhysRevB.69.054403} {\bibfield
  {journal} {\bibinfo  {journal} {Phys. Rev. B}\ }\textbf {\bibinfo {volume}
  {69}},\ \bibinfo {pages} {054403} (\bibinfo {year} {2004})}\BibitemShut
  {NoStop}%
\bibitem [{\citenamefont {Kolovsky}\ and\ \citenamefont
  {Buchleitner}(2004)}]{Kolovsky_Buchleitner_EPL2004}%
  \BibitemOpen
  \bibfield  {author} {\bibinfo {author} {\bibfnamefont {A.~R.}\ \bibnamefont
  {Kolovsky}}\ and\ \bibinfo {author} {\bibfnamefont {A.}~\bibnamefont
  {Buchleitner}},\ }\href {http://stacks.iop.org/0295-5075/68/i=5/a=632}
  {\bibfield  {journal} {\bibinfo  {journal} {Europhys. Lett.}\ }\textbf
  {\bibinfo {volume} {68}},\ \bibinfo {pages} {632} (\bibinfo {year}
  {2004})}\BibitemShut {NoStop}%
\bibitem [{\citenamefont {Kudo}\ and\ \citenamefont
  {Deguchi}(2005)}]{KudoDeguchi_JPSJ2005}%
  \BibitemOpen
  \bibfield  {author} {\bibinfo {author} {\bibfnamefont {K.}~\bibnamefont
  {Kudo}}\ and\ \bibinfo {author} {\bibfnamefont {T.}~\bibnamefont {Deguchi}},\
  }\href {\doibase 10.1143/JPSJ.74.1992} {\bibfield  {journal} {\bibinfo
  {journal} {J. Phys. Soc. Jpn}\ }\textbf {\bibinfo {volume} {74}},\ \bibinfo
  {pages} {1992} (\bibinfo {year} {2005})}\BibitemShut {NoStop}%
\bibitem [{\citenamefont {Karthik}\ \emph {et~al.}(2007)\citenamefont
  {Karthik}, \citenamefont {Sharma},\ and\ \citenamefont
  {Lakshminarayan}}]{KarthikEA_PRA2007}%
  \BibitemOpen
  \bibfield  {author} {\bibinfo {author} {\bibfnamefont {J.}~\bibnamefont
  {Karthik}}, \bibinfo {author} {\bibfnamefont {A.}~\bibnamefont {Sharma}}, \
  and\ \bibinfo {author} {\bibfnamefont {A.}~\bibnamefont {Lakshminarayan}},\
  }\href {\doibase 10.1103/PhysRevA.75.022304} {\bibfield  {journal} {\bibinfo
  {journal} {Phys. Rev. A}\ }\textbf {\bibinfo {volume} {75}},\ \bibinfo
  {pages} {022304} (\bibinfo {year} {2007})}\BibitemShut {NoStop}%
\bibitem [{\citenamefont {Santos}(2009)}]{Santos_JMathPhys2009}%
  \BibitemOpen
  \bibfield  {author} {\bibinfo {author} {\bibfnamefont {L.~F.}\ \bibnamefont
  {Santos}},\ }\href {\doibase 10.1063/1.3181223} {\bibfield  {journal}
  {\bibinfo  {journal} {J. Math. Phys.}\ }\textbf {\bibinfo {volume} {50}},\
  \bibinfo {pages} {095211} (\bibinfo {year} {2009})}\BibitemShut {NoStop}%
\bibitem [{\citenamefont {Kollath}\ \emph {et~al.}()\citenamefont {Kollath},
  \citenamefont {Roux}, \citenamefont {Biroli},\ and\ \citenamefont
  {L{\"a}uchli}}]{KollathEA2010}%
  \BibitemOpen
  \bibfield  {author} {\bibinfo {author} {\bibfnamefont {C.}~\bibnamefont
  {Kollath}}, \bibinfo {author} {\bibfnamefont {G.}~\bibnamefont {Roux}},
  \bibinfo {author} {\bibfnamefont {G.}~\bibnamefont {Biroli}}, \ and\ \bibinfo
  {author} {\bibfnamefont {A.~M.}\ \bibnamefont {L{\"a}uchli}},\ }\href
  {http://stacks.iop.org/1742-5468/2010/i=08/a=P08011} {\bibfield  {journal}
  {\bibinfo  {journal} {J. Stat. Mech.}\ }\textbf {\bibinfo {volume} {2010}},\
  \bibinfo {pages} {P08011}}\BibitemShut {NoStop}%
\bibitem [{\citenamefont {Santos}\ and\ \citenamefont
  {Rigol}(2010{\natexlab{a}})}]{SantosRigol2010a}%
  \BibitemOpen
  \bibfield  {author} {\bibinfo {author} {\bibfnamefont {L.~F.}\ \bibnamefont
  {Santos}}\ and\ \bibinfo {author} {\bibfnamefont {M.}~\bibnamefont {Rigol}},\
  }\href {\doibase 10.1103/PhysRevE.81.036206} {\bibfield  {journal} {\bibinfo
  {journal} {Phys. Rev. E}\ }\textbf {\bibinfo {volume} {81}},\ \bibinfo
  {pages} {036206} (\bibinfo {year} {2010}{\natexlab{a}})}\BibitemShut
  {NoStop}%
\bibitem [{\citenamefont {Santos}\ \emph {et~al.}(2012)\citenamefont {Santos},
  \citenamefont {Borgonovi},\ and\ \citenamefont {Izrailev}}]{SantosEA2012}%
  \BibitemOpen
  \bibfield  {author} {\bibinfo {author} {\bibfnamefont {L.~F.}\ \bibnamefont
  {Santos}}, \bibinfo {author} {\bibfnamefont {F.}~\bibnamefont {Borgonovi}}, \
  and\ \bibinfo {author} {\bibfnamefont {F.~M.}\ \bibnamefont {Izrailev}},\
  }\href {\doibase 10.1103/PhysRevE.85.036209} {\bibfield  {journal} {\bibinfo
  {journal} {Phys. Rev. E}\ }\textbf {\bibinfo {volume} {85}},\ \bibinfo
  {pages} {036209} (\bibinfo {year} {2012})}\BibitemShut {NoStop}%
\bibitem [{\citenamefont {Atas}\ \emph {et~al.}(2013)\citenamefont {Atas},
  \citenamefont {Bogomolny}, \citenamefont {Giraud},\ and\ \citenamefont
  {Roux}}]{BogomolnyRoux_PRL2014}%
  \BibitemOpen
  \bibfield  {author} {\bibinfo {author} {\bibfnamefont {Y.~Y.}\ \bibnamefont
  {Atas}}, \bibinfo {author} {\bibfnamefont {E.}~\bibnamefont {Bogomolny}},
  \bibinfo {author} {\bibfnamefont {O.}~\bibnamefont {Giraud}}, \ and\ \bibinfo
  {author} {\bibfnamefont {G.}~\bibnamefont {Roux}},\ }\href {\doibase
  10.1103/PhysRevLett.110.084101} {\bibfield  {journal} {\bibinfo  {journal}
  {Phys. Rev. Lett.}\ }\textbf {\bibinfo {volume} {110}},\ \bibinfo {pages}
  {084101} (\bibinfo {year} {2013})}\BibitemShut {NoStop}%
\bibitem [{\citenamefont {{B{\"a}cker}}\ and\ \citenamefont
  {{Schubert}}(2002)}]{BaeckerSchubert_JPA2002}%
  \BibitemOpen
  \bibfield  {author} {\bibinfo {author} {\bibfnamefont {A.}~\bibnamefont
  {{B{\"a}cker}}}\ and\ \bibinfo {author} {\bibfnamefont {R.}~\bibnamefont
  {{Schubert}}},\ }\href@noop {} {\bibfield  {journal} {\bibinfo  {journal} {J.
  Phys. A: Math. Gen.}\ }\textbf {\bibinfo {volume} {35}},\ \bibinfo {pages}
  {527} (\bibinfo {year} {2002})}\BibitemShut {NoStop}%
\bibitem [{\citenamefont {Heller}(1984)}]{Hel1984}%
  \BibitemOpen
  \bibfield  {author} {\bibinfo {author} {\bibfnamefont {E.~J.}\ \bibnamefont
  {Heller}},\ }\href {\doibase 10.1103/PhysRevLett.53.1515} {\bibfield
  {journal} {\bibinfo  {journal} {Phys. Rev. Lett.}\ }\textbf {\bibinfo
  {volume} {53}},\ \bibinfo {pages} {1515} (\bibinfo {year}
  {1984})}\BibitemShut {NoStop}%
\bibitem [{\citenamefont {B\"acker}\ \emph {et~al.}(1997)\citenamefont
  {B\"acker}, \citenamefont {Schubert},\ and\ \citenamefont
  {Stifter}}]{BaeSchSti1997}%
  \BibitemOpen
  \bibfield  {author} {\bibinfo {author} {\bibfnamefont {A.}~\bibnamefont
  {B\"acker}}, \bibinfo {author} {\bibfnamefont {R.}~\bibnamefont {Schubert}},
  \ and\ \bibinfo {author} {\bibfnamefont {P.}~\bibnamefont {Stifter}},\ }\href
  {\doibase 10.1088/0305-4470/30/19/017} {\bibfield  {journal} {\bibinfo
  {journal} {J. Phys. A: Math. Gen.}\ }\textbf {\bibinfo {volume} {30}},\
  \bibinfo {pages} {6783} (\bibinfo {year} {1997})}\BibitemShut {NoStop}%
\bibitem [{\citenamefont {Berry}(1977)}]{Ber1977b}%
  \BibitemOpen
  \bibfield  {author} {\bibinfo {author} {\bibfnamefont {M.~V.}\ \bibnamefont
  {Berry}},\ }\href {\doibase 10.1088/0305-4470/10/12/016} {\bibfield
  {journal} {\bibinfo  {journal} {J. Phys. A: Math. Gen.}\ }\textbf {\bibinfo
  {volume} {10}},\ \bibinfo {pages} {2083} (\bibinfo {year}
  {1977})}\BibitemShut {NoStop}%
\bibitem [{\citenamefont {Porter}\ and\ \citenamefont
  {Thomas}(1956)}]{PorterThomas_PR1956}%
  \BibitemOpen
  \bibfield  {author} {\bibinfo {author} {\bibfnamefont {C.~E.}\ \bibnamefont
  {Porter}}\ and\ \bibinfo {author} {\bibfnamefont {R.~G.}\ \bibnamefont
  {Thomas}},\ }\href {\doibase 10.1103/PhysRev.104.483} {\bibfield  {journal}
  {\bibinfo  {journal} {Phys. Rev.}\ }\textbf {\bibinfo {volume} {104}},\
  \bibinfo {pages} {483} (\bibinfo {year} {1956})}\BibitemShut {NoStop}%
\bibitem [{\citenamefont {Luitz}\ and\ \citenamefont
  {Bar~Lev}(2016)}]{LuitzBarlev_PRL16}%
  \BibitemOpen
  \bibfield  {author} {\bibinfo {author} {\bibfnamefont {D.~J.}\ \bibnamefont
  {Luitz}}\ and\ \bibinfo {author} {\bibfnamefont {Y.}~\bibnamefont
  {Bar~Lev}},\ }\href {\doibase 10.1103/PhysRevLett.117.170404} {\bibfield
  {journal} {\bibinfo  {journal} {Phys. Rev. Lett.}\ }\textbf {\bibinfo
  {volume} {117}},\ \bibinfo {pages} {170404} (\bibinfo {year}
  {2016})}\BibitemShut {NoStop}%
\bibitem [{\citenamefont {Mondaini}\ and\ \citenamefont
  {Rigol}(2017)}]{MondainiRigol_PRE2017}%
  \BibitemOpen
  \bibfield  {author} {\bibinfo {author} {\bibfnamefont {R.}~\bibnamefont
  {Mondaini}}\ and\ \bibinfo {author} {\bibfnamefont {M.}~\bibnamefont
  {Rigol}},\ }\href {\doibase 10.1103/PhysRevE.96.012157} {\bibfield  {journal}
  {\bibinfo  {journal} {Phys. Rev. E}\ }\textbf {\bibinfo {volume} {96}},\
  \bibinfo {pages} {012157} (\bibinfo {year} {2017})}\BibitemShut {NoStop}%
\bibitem [{\citenamefont {Mondaini}\ \emph {et~al.}(2016)\citenamefont
  {Mondaini}, \citenamefont {Fratus}, \citenamefont {Srednicki},\ and\
  \citenamefont {Rigol}}]{MondainiSrednickiRigol_PRE16}%
  \BibitemOpen
  \bibfield  {author} {\bibinfo {author} {\bibfnamefont {R.}~\bibnamefont
  {Mondaini}}, \bibinfo {author} {\bibfnamefont {K.~R.}\ \bibnamefont
  {Fratus}}, \bibinfo {author} {\bibfnamefont {M.}~\bibnamefont {Srednicki}}, \
  and\ \bibinfo {author} {\bibfnamefont {M.}~\bibnamefont {Rigol}},\ }\href
  {\doibase 10.1103/PhysRevE.93.032104} {\bibfield  {journal} {\bibinfo
  {journal} {Phys. Rev. E}\ }\textbf {\bibinfo {volume} {93}},\ \bibinfo
  {pages} {032104} (\bibinfo {year} {2016})}\BibitemShut {NoStop}%
\bibitem [{\citenamefont {Atas}\ and\ \citenamefont
  {Bogomolny}(2017)}]{AtasBogomolny_JPhysA2017}%
  \BibitemOpen
  \bibfield  {author} {\bibinfo {author} {\bibfnamefont {Y.~Y.}\ \bibnamefont
  {Atas}}\ and\ \bibinfo {author} {\bibfnamefont {E.}~\bibnamefont
  {Bogomolny}},\ }\href {http://stacks.iop.org/1751-8121/50/i=38/a=385102}
  {\bibfield  {journal} {\bibinfo  {journal} {J. Phys. A: Math. Gen.}\ }\textbf
  {\bibinfo {volume} {50}},\ \bibinfo {pages} {385102} (\bibinfo {year}
  {2017})}\BibitemShut {NoStop}%
\bibitem [{\citenamefont {Kullback}\ and\ \citenamefont
  {Leibler}(1951)}]{KullbackLeibler1951}%
  \BibitemOpen
  \bibfield  {author} {\bibinfo {author} {\bibfnamefont {S.}~\bibnamefont
  {Kullback}}\ and\ \bibinfo {author} {\bibfnamefont {R.~A.}\ \bibnamefont
  {Leibler}},\ }\href {\doibase 10.1214/aoms/1177729694} {\bibfield  {journal}
  {\bibinfo  {journal} {Ann. Math. Statist.}\ }\textbf {\bibinfo {volume}
  {22}},\ \bibinfo {pages} {79} (\bibinfo {year} {1951})}\BibitemShut {NoStop}%
\bibitem [{Note1()}]{Note1}%
  \BibitemOpen
  \bibinfo {note} {Here, we separate the system in two spatial parts A and B of
  (nearly) equal size. Given a state $\protect {\delimiter 69640972 \psi
  \delimiter "526930B }$, the entanglement entropy is $S=-\mathop {\protect
  \mathrm {Tr}}\nolimits \rho _\protect \mathrm {A}\protect \qopname \relax
  o{log}\rho _\protect \mathrm {A}$, where $\rho _\protect \mathrm {A}=\mathop
  {\mathop {\protect \mathrm {Tr}}\nolimits _\protect \mathrm {B}}\nolimits
  \protect {\delimiter 69640972 \psi \delimiter "526930B }\protect {\delimiter
  "426830A \psi \delimiter 86418188 }$ is the reduced density matrix. The
  spatial partition is most relevant to the present study of amplitudes in the
  basis of real-space configurations.}\BibitemShut {Stop}%
\bibitem [{\citenamefont {Beugeling}\ \emph {et~al.}()\citenamefont
  {Beugeling}, \citenamefont {Andreanov},\ and\ \citenamefont
  {Haque}}]{BeugelingEA2015JStatMech}%
  \BibitemOpen
  \bibfield  {author} {\bibinfo {author} {\bibfnamefont {W.}~\bibnamefont
  {Beugeling}}, \bibinfo {author} {\bibfnamefont {A.}~\bibnamefont
  {Andreanov}}, \ and\ \bibinfo {author} {\bibfnamefont {M.}~\bibnamefont
  {Haque}},\ }\href {\doibase 10.1088/1742-5468/2015/02/P02002} {\bibfield
  {journal} {\bibinfo  {journal} {J. Stat. Mech.}\ }\textbf {\bibinfo {volume}
  {2015}},\ \bibinfo {pages} {P02002}}\BibitemShut {NoStop}%
\bibitem [{\citenamefont {Alba}\ \emph {et~al.}()\citenamefont {Alba},
  \citenamefont {Fagotti},\ and\ \citenamefont
  {Calabrese}}]{alba2009entanglement}%
  \BibitemOpen
  \bibfield  {author} {\bibinfo {author} {\bibfnamefont {V.}~\bibnamefont
  {Alba}}, \bibinfo {author} {\bibfnamefont {M.}~\bibnamefont {Fagotti}}, \
  and\ \bibinfo {author} {\bibfnamefont {P.}~\bibnamefont {Calabrese}},\ }\href
  {http://stacks.iop.org/1742-5468/2009/i=10/a=P10020} {\bibfield  {journal}
  {\bibinfo  {journal} {J. Stat. Mech.}\ }\textbf {\bibinfo {volume} {2009}},\
  \bibinfo {pages} {P10020}}\BibitemShut {NoStop}%
\bibitem [{\citenamefont {Garrison}\ and\ \citenamefont
  {Grover}(2018)}]{GarrisonGrover2018PRX}%
  \BibitemOpen
  \bibfield  {author} {\bibinfo {author} {\bibfnamefont {J.~R.}\ \bibnamefont
  {Garrison}}\ and\ \bibinfo {author} {\bibfnamefont {T.}~\bibnamefont
  {Grover}},\ }\href {\doibase 10.1103/PhysRevX.8.021026} {\bibfield  {journal}
  {\bibinfo  {journal} {Phys. Rev. X}\ }\textbf {\bibinfo {volume} {8}},\
  \bibinfo {pages} {021026} (\bibinfo {year} {2018})}\BibitemShut {NoStop}%
\bibitem [{\citenamefont {Vidmar}\ and\ \citenamefont
  {Rigol}(2017)}]{VidmarRigol2017_entanglement}%
  \BibitemOpen
  \bibfield  {author} {\bibinfo {author} {\bibfnamefont {L.}~\bibnamefont
  {Vidmar}}\ and\ \bibinfo {author} {\bibfnamefont {M.}~\bibnamefont {Rigol}},\
  }\href {\doibase 10.1103/PhysRevLett.119.220603} {\bibfield  {journal}
  {\bibinfo  {journal} {Phys. Rev. Lett.}\ }\textbf {\bibinfo {volume} {119}},\
  \bibinfo {pages} {220603} (\bibinfo {year} {2017})}\BibitemShut {NoStop}%
\bibitem [{\citenamefont {Santos}\ and\ \citenamefont
  {Rigol}(2010{\natexlab{b}})}]{SantosRigol2010}%
  \BibitemOpen
  \bibfield  {author} {\bibinfo {author} {\bibfnamefont {L.~F.}\ \bibnamefont
  {Santos}}\ and\ \bibinfo {author} {\bibfnamefont {M.}~\bibnamefont {Rigol}},\
  }\href {\doibase 10.1103/PhysRevE.82.031130} {\bibfield  {journal} {\bibinfo
  {journal} {Phys. Rev. E}\ }\textbf {\bibinfo {volume} {82}},\ \bibinfo
  {pages} {031130} (\bibinfo {year} {2010}{\natexlab{b}})}\BibitemShut
  {NoStop}%
\bibitem [{\citenamefont {Tao}\ and\ \citenamefont {Vu}(2012)}]{TaoVu2012}%
  \BibitemOpen
  \bibfield  {author} {\bibinfo {author} {\bibfnamefont {T.}~\bibnamefont
  {Tao}}\ and\ \bibinfo {author} {\bibfnamefont {V.}~\bibnamefont {Vu}},\
  }\href {\doibase http://dx.doi.org/10.1016/j.aim.2012.05.006} {\bibfield
  {journal} {\bibinfo  {journal} {Adv. Math.}\ }\textbf {\bibinfo {volume}
  {231}},\ \bibinfo {pages} {74 } (\bibinfo {year} {2012})}\BibitemShut
  {NoStop}%
\bibitem [{\citenamefont {Nguyen}\ and\ \citenamefont
  {Vu}(2014)}]{NguyenVu2014}%
  \BibitemOpen
  \bibfield  {author} {\bibinfo {author} {\bibfnamefont {H.~H.}\ \bibnamefont
  {Nguyen}}\ and\ \bibinfo {author} {\bibfnamefont {V.}~\bibnamefont {Vu}},\
  }\href {\doibase 10.1214/12-AOP791} {\bibfield  {journal} {\bibinfo
  {journal} {Ann. Probab.}\ }\textbf {\bibinfo {volume} {42}},\ \bibinfo
  {pages} {146} (\bibinfo {year} {2014})}\BibitemShut {NoStop}%
\bibitem [{\citenamefont {Neuenhahn}\ and\ \citenamefont
  {Marquardt}(2012)}]{NeuenhahnMarquardt2012}%
  \BibitemOpen
  \bibfield  {author} {\bibinfo {author} {\bibfnamefont {C.}~\bibnamefont
  {Neuenhahn}}\ and\ \bibinfo {author} {\bibfnamefont {F.}~\bibnamefont
  {Marquardt}},\ }\href {\doibase 10.1103/PhysRevE.85.060101} {\bibfield
  {journal} {\bibinfo  {journal} {Phys. Rev. E}\ }\textbf {\bibinfo {volume}
  {85}},\ \bibinfo {pages} {060101} (\bibinfo {year} {2012})}\BibitemShut
  {NoStop}%
\bibitem [{\citenamefont {Beugeling}\ \emph {et~al.}(2014)\citenamefont
  {Beugeling}, \citenamefont {Moessner},\ and\ \citenamefont
  {Haque}}]{BeugelingEA2014PRE}%
  \BibitemOpen
  \bibfield  {author} {\bibinfo {author} {\bibfnamefont {W.}~\bibnamefont
  {Beugeling}}, \bibinfo {author} {\bibfnamefont {R.}~\bibnamefont {Moessner}},
  \ and\ \bibinfo {author} {\bibfnamefont {M.}~\bibnamefont {Haque}},\ }\href
  {\doibase 10.1103/PhysRevE.89.042112} {\bibfield  {journal} {\bibinfo
  {journal} {Phys. Rev. E}\ }\textbf {\bibinfo {volume} {89}},\ \bibinfo
  {pages} {042112} (\bibinfo {year} {2014})}\BibitemShut {NoStop}%
\bibitem [{\citenamefont {Beugeling}\ \emph {et~al.}(2015)\citenamefont
  {Beugeling}, \citenamefont {Moessner},\ and\ \citenamefont
  {Haque}}]{BeugelingEA2015PRE}%
  \BibitemOpen
  \bibfield  {author} {\bibinfo {author} {\bibfnamefont {W.}~\bibnamefont
  {Beugeling}}, \bibinfo {author} {\bibfnamefont {R.}~\bibnamefont {Moessner}},
  \ and\ \bibinfo {author} {\bibfnamefont {M.}~\bibnamefont {Haque}},\ }\href
  {\doibase 10.1103/PhysRevE.91.012144} {\bibfield  {journal} {\bibinfo
  {journal} {Phys. Rev. E}\ }\textbf {\bibinfo {volume} {91}},\ \bibinfo
  {pages} {012144} (\bibinfo {year} {2015})}\BibitemShut {NoStop}%
\bibitem [{\citenamefont {Samajdar}\ and\ \citenamefont
  {Jain}(2018)}]{SamajdarJain_integrablebilliards}%
  \BibitemOpen
  \bibfield  {author} {\bibinfo {author} {\bibfnamefont {R.}~\bibnamefont
  {Samajdar}}\ and\ \bibinfo {author} {\bibfnamefont {S.~R.}\ \bibnamefont
  {Jain}},\ }\href {\doibase 10.1063/1.5006320} {\bibfield  {journal} {\bibinfo
   {journal} {J. Math. Phys.}\ }\textbf {\bibinfo {volume} {59}},\ \bibinfo
  {pages} {012103} (\bibinfo {year} {2018})}\BibitemShut {NoStop}%
\bibitem [{\citenamefont {Horoi}\ \emph {et~al.}(1995)\citenamefont {Horoi},
  \citenamefont {Zelevinsky},\ and\ \citenamefont
  {Brown}}]{Horoi_Zelevinsky_PRL1995}%
  \BibitemOpen
  \bibfield  {author} {\bibinfo {author} {\bibfnamefont {M.}~\bibnamefont
  {Horoi}}, \bibinfo {author} {\bibfnamefont {V.}~\bibnamefont {Zelevinsky}}, \
  and\ \bibinfo {author} {\bibfnamefont {B.~A.}\ \bibnamefont {Brown}},\ }\href
  {\doibase 10.1103/PhysRevLett.74.5194} {\bibfield  {journal} {\bibinfo
  {journal} {Phys. Rev. Lett.}\ }\textbf {\bibinfo {volume} {74}},\ \bibinfo
  {pages} {5194} (\bibinfo {year} {1995})}\BibitemShut {NoStop}%
\bibitem [{\citenamefont {Zelevinsky}\ \emph {et~al.}(1996)\citenamefont
  {Zelevinsky}, \citenamefont {Brown}, \citenamefont {Frazier},\ and\
  \citenamefont {Horoi}}]{ZelevinskyEA_PhysRep1996}%
  \BibitemOpen
  \bibfield  {author} {\bibinfo {author} {\bibfnamefont {V.}~\bibnamefont
  {Zelevinsky}}, \bibinfo {author} {\bibfnamefont {B.~A.}\ \bibnamefont
  {Brown}}, \bibinfo {author} {\bibfnamefont {N.}~\bibnamefont {Frazier}}, \
  and\ \bibinfo {author} {\bibfnamefont {M.}~\bibnamefont {Horoi}},\ }\href
  {\doibase 10.1016/S0370-1573(96)00007-5} {\bibfield  {journal} {\bibinfo
  {journal} {Phys. Rep.}\ }\textbf {\bibinfo {volume} {276}},\ \bibinfo {pages}
  {85} (\bibinfo {year} {1996})}\BibitemShut {NoStop}%
\bibitem [{\citenamefont {Flambaum}\ \emph {et~al.}(1996)\citenamefont
  {Flambaum}, \citenamefont {Izrailev},\ and\ \citenamefont
  {Casati}}]{Flambaum_Izrailev_Casati_PRE1996}%
  \BibitemOpen
  \bibfield  {author} {\bibinfo {author} {\bibfnamefont {V.~V.}\ \bibnamefont
  {Flambaum}}, \bibinfo {author} {\bibfnamefont {F.~M.}\ \bibnamefont
  {Izrailev}}, \ and\ \bibinfo {author} {\bibfnamefont {G.}~\bibnamefont
  {Casati}},\ }\href {\doibase 10.1103/PhysRevE.54.2136} {\bibfield  {journal}
  {\bibinfo  {journal} {Phys. Rev. E}\ }\textbf {\bibinfo {volume} {54}},\
  \bibinfo {pages} {2136} (\bibinfo {year} {1996})}\BibitemShut {NoStop}%
\bibitem [{\citenamefont {Flambaum}\ and\ \citenamefont
  {Izrailev}(1997)}]{Flambaum_Izrailev_PRE1997}%
  \BibitemOpen
  \bibfield  {author} {\bibinfo {author} {\bibfnamefont {V.~V.}\ \bibnamefont
  {Flambaum}}\ and\ \bibinfo {author} {\bibfnamefont {F.~M.}\ \bibnamefont
  {Izrailev}},\ }\href {\doibase 10.1103/PhysRevE.56.5144} {\bibfield
  {journal} {\bibinfo  {journal} {Phys. Rev. E}\ }\textbf {\bibinfo {volume}
  {56}},\ \bibinfo {pages} {5144} (\bibinfo {year} {1997})}\BibitemShut
  {NoStop}%
\bibitem [{\citenamefont {Guhr}\ \emph {et~al.}(1998)\citenamefont {Guhr},
  \citenamefont {M{\"u}ller-Groeling},\ and\ \citenamefont
  {Weidenm{\"u}ller}}]{Guhr_Weidemuller_PhysRep1998}%
  \BibitemOpen
  \bibfield  {author} {\bibinfo {author} {\bibfnamefont {T.}~\bibnamefont
  {Guhr}}, \bibinfo {author} {\bibfnamefont {A.}~\bibnamefont
  {M{\"u}ller-Groeling}}, \ and\ \bibinfo {author} {\bibfnamefont {H.~A.}\
  \bibnamefont {Weidenm{\"u}ller}},\ }\href {\doibase
  http://dx.doi.org/10.1016/S0370-1573(97)00088-4} {\bibfield  {journal}
  {\bibinfo  {journal} {Phys. Rep.}\ }\textbf {\bibinfo {volume} {299}},\
  \bibinfo {pages} {189} (\bibinfo {year} {1998})}\BibitemShut {NoStop}%
\bibitem [{\citenamefont {Rigol}\ and\ \citenamefont
  {Santos}(2010)}]{RigolSantos2010}%
  \BibitemOpen
  \bibfield  {author} {\bibinfo {author} {\bibfnamefont {M.}~\bibnamefont
  {Rigol}}\ and\ \bibinfo {author} {\bibfnamefont {L.~F.}\ \bibnamefont
  {Santos}},\ }\href {\doibase 10.1103/PhysRevA.82.011604} {\bibfield
  {journal} {\bibinfo  {journal} {Phys. Rev. A}\ }\textbf {\bibinfo {volume}
  {82}},\ \bibinfo {pages} {011604} (\bibinfo {year} {2010})}\BibitemShut
  {NoStop}%
\bibitem [{\citenamefont {{Nandkishore}}\ and\ \citenamefont
  {{Huse}}(2015)}]{NandkishoreHuse_AnnuRev2015}%
  \BibitemOpen
  \bibfield  {author} {\bibinfo {author} {\bibfnamefont {R.}~\bibnamefont
  {{Nandkishore}}}\ and\ \bibinfo {author} {\bibfnamefont {D.~A.}\ \bibnamefont
  {{Huse}}},\ }\href {\doibase 10.1146/annurev-conmatphys-031214-014726}
  {\bibfield  {journal} {\bibinfo  {journal} {Annu. Rev. Condens. Matter
  Phys.}\ }\textbf {\bibinfo {volume} {6}},\ \bibinfo {pages} {15} (\bibinfo
  {year} {2015})}\BibitemShut {NoStop}%
\bibitem [{\citenamefont {{Altman}}\ and\ \citenamefont
  {{Vosk}}(2015)}]{AltmanVosk_AnnuRev2015}%
  \BibitemOpen
  \bibfield  {author} {\bibinfo {author} {\bibfnamefont {E.}~\bibnamefont
  {{Altman}}}\ and\ \bibinfo {author} {\bibfnamefont {R.}~\bibnamefont
  {{Vosk}}},\ }\href {\doibase 10.1146/annurev-conmatphys-031214-014701}
  {\bibfield  {journal} {\bibinfo  {journal} {Annu. Rev. Condens. Matter
  Phys.}\ }\textbf {\bibinfo {volume} {6}},\ \bibinfo {pages} {383} (\bibinfo
  {year} {2015})}\BibitemShut {NoStop}%
\bibitem [{\citenamefont {Armstrong}\ \emph {et~al.}(2012)\citenamefont
  {Armstrong}, \citenamefont {\AA{}berg}, \citenamefont {Reimann},\ and\
  \citenamefont {Zelevinsky}}]{ArmstrongZelevinsky_PRE2012}%
  \BibitemOpen
  \bibfield  {author} {\bibinfo {author} {\bibfnamefont {J.~R.}\ \bibnamefont
  {Armstrong}}, \bibinfo {author} {\bibfnamefont {S.}~\bibnamefont
  {\AA{}berg}}, \bibinfo {author} {\bibfnamefont {S.~M.}\ \bibnamefont
  {Reimann}}, \ and\ \bibinfo {author} {\bibfnamefont {V.~G.}\ \bibnamefont
  {Zelevinsky}},\ }\href {\doibase 10.1103/PhysRevE.86.066204} {\bibfield
  {journal} {\bibinfo  {journal} {Phys. Rev. E}\ }\textbf {\bibinfo {volume}
  {86}},\ \bibinfo {pages} {066204} (\bibinfo {year} {2012})}\BibitemShut
  {NoStop}%
\bibitem [{\citenamefont {Haque}\ and\ \citenamefont
  {McClarty}(2017)}]{HaqueMcClarty_SYKETH}%
  \BibitemOpen
  \bibfield  {author} {\bibinfo {author} {\bibfnamefont {M.}~\bibnamefont
  {Haque}}\ and\ \bibinfo {author} {\bibfnamefont {P.}~\bibnamefont
  {McClarty}},\ }\href@noop {} {\bibfield  {journal} {\bibinfo  {journal}
  {ArXiv e-prints}\ } (\bibinfo {year} {2017})},\ \Eprint
  {http://arxiv.org/abs/1711.02360} {arXiv:1711.02360 [cond-mat.stat-mech]}
  \BibitemShut {NoStop}%
\bibitem [{\citenamefont {Atas}\ and\ \citenamefont
  {Bogomolny}(2012)}]{AtasBogomolny_PRE12}%
  \BibitemOpen
  \bibfield  {author} {\bibinfo {author} {\bibfnamefont {Y.~Y.}\ \bibnamefont
  {Atas}}\ and\ \bibinfo {author} {\bibfnamefont {E.}~\bibnamefont
  {Bogomolny}},\ }\href {\doibase 10.1103/PhysRevE.86.021104} {\bibfield
  {journal} {\bibinfo  {journal} {Phys. Rev. E}\ }\textbf {\bibinfo {volume}
  {86}},\ \bibinfo {pages} {021104} (\bibinfo {year} {2012})}\BibitemShut
  {NoStop}%
\bibitem [{\citenamefont {Luitz}\ \emph {et~al.}(2014)\citenamefont {Luitz},
  \citenamefont {Alet},\ and\ \citenamefont
  {Laflorencie}}]{LuitzAletLaflorencie_PRL14}%
  \BibitemOpen
  \bibfield  {author} {\bibinfo {author} {\bibfnamefont {D.~J.}\ \bibnamefont
  {Luitz}}, \bibinfo {author} {\bibfnamefont {F.}~\bibnamefont {Alet}}, \ and\
  \bibinfo {author} {\bibfnamefont {N.}~\bibnamefont {Laflorencie}},\ }\href
  {\doibase 10.1103/PhysRevLett.112.057203} {\bibfield  {journal} {\bibinfo
  {journal} {Phys. Rev. Lett.}\ }\textbf {\bibinfo {volume} {112}},\ \bibinfo
  {pages} {057203} (\bibinfo {year} {2014})}\BibitemShut {NoStop}%
\bibitem [{\citenamefont {Abramowitz}\ and\ \citenamefont
  {Stegun}(1972)}]{AbramowitzStegun1972}%
  \BibitemOpen
  \bibfield  {author} {\bibinfo {author} {\bibfnamefont {M.}~\bibnamefont
  {Abramowitz}}\ and\ \bibinfo {author} {\bibfnamefont {I.~A.}\ \bibnamefont
  {Stegun}},\ }\href@noop {} {\emph {\bibinfo {title} {Handbook of Mathematical
  Functions}}},\ Dover Books on Mathematics\ (\bibinfo  {publisher} {Dover
  Publications},\ \bibinfo {year} {1972})\BibitemShut {NoStop}%
\end{thebibliography}
%

\end{document}